\documentclass[prb,twocolumn,groupedaddress,shownopacs,amssymb]{revtex4-2}
\usepackage{graphicx,psfrag,amsmath,amssymb}

\usepackage[usenames, dvipsnames]{color}

\usepackage{hyperref} 
\hypersetup{
    colorlinks=true,
    linkcolor=Blue,
    citecolor=Blue,
    filecolor=Blue,      
    urlcolor=Blue,
    pdftitle={Disordered BCS-BEC near Tc},
    }

\begin{document}

\title{Self-consistent inclusion of disorder in the BCS-BEC 
crossover near the critical temperature
}

\author{M. Iskin}
\affiliation{
Department of Physics, Ko\c{c} University, Rumelifeneri Yolu, 
34450 Sar\i yer, Istanbul, T\"urkiye
}

\date{\today}

\begin{abstract}

We develop a systematic theoretical approach to incorporate the 
effects of a static white-noise disorder into the BCS-BEC 
crossover near the critical temperature ($T_c$) of the superfluid 
transition. Starting from a functional-integral formulation in 
momentum-frequency space, we derive an effective thermodynamic 
potential that fully accounts for Gaussian fluctuations of the 
order-parameter field and its coupling to the disorder potential. 
The effective action, expanded to second order in both the disorder 
potential and the bosonic field, naturally involves third- and 
fourth-order terms arising from the logarithmic expansion near $T_c$. 
By providing a controlled description of pairing fluctuations and 
disorder effects, this formalism correctly recovers the 
well-established BCS and BEC limits. This ensures a consistent 
physical foundation for analyzing the entire BCS-BEC crossover, 
effectively anchoring the intermediate regime between these two 
analytically robust endpoints.
The approach applies equally to continuum and lattice systems and 
provides a natural framework for generalizations to multiband models.

\end{abstract}
\maketitle

\section{Introduction}
\label{sec:intro}

The study of disordered quantum gases has emerged as a powerful  
setting for exploring the interplay among interactions, coherence  
and localization in highly controllable environments  
~\cite{ospelkaus06, modugno10, deissler10, sanchez10, kondov11,  
jendrzejewski12, krinner13, krinner15, choi16, cappellaro19, nagler20,  
koch24, russ25}.  
Ultracold Fermi gases, in particular, provide an ideal platform  
for realizing the BCS-BEC crossover under tunable  
conditions~\cite{giorgini08, ketterle08, strinati18, hartke25},  
where both the interaction strength and the disorder potential  
can be precisely controlled via Feshbach resonances and optical  
speckle fields.
While the crossover is well characterized in clean systems, the  
combined effects of disorder and strong pairing correlations remain  
only partially understood~\cite{sanchez10, nagler20, koch24}.  
Theoretically, Fermi gases interpolate between two qualitatively  
distinct limits: on the BCS side, Anderson’s theorem guarantees the  
robustness of $T_c$ against weak nonmagnetic disorder, whereas  
on the BEC side, the transition is governed by phase coherence among  
tightly-bound fermion pairs, which is highly susceptible to localization  
and incoherence~\cite{orso07, han11, palestini13, iskin25c}.  
A unified finite-temperature description valid throughout the  
BCS-BEC crossover is therefore crucial for interpreting cold-atom  
experiments and for connecting continuum and lattice models of  
disordered superfluidity.  
Previous theoretical studies have either focused on the  
zero-temperature ($T = 0$) superfluid phase~\cite{orso07, iskin25c}  
or determined $T_c$ through semiclassical analyses performed within  
the local-density approximation and the replica trick~\cite{han11}.  
Other studies have employed diagrammatic methods at the lowest  
nontrivial order for the normal phase above $T_c$, complemented by  
accurate numerical evaluations of the corresponding diagrammatic  
contributions~\cite{palestini13}, or have relied on $T$-matrix 
approaches~\cite{che17}. A fully self-consistent treatment valid 
across the entire crossover region at finite temperatures is still lacking.  

Building on a functional-integral framework in momentum-frequency  
space, we construct an effective thermodynamic potential that  
incorporates Gaussian fluctuations of the order parameter together  
with their coupling to a static disorder potential. Crucially, our  
formulation provides a fully quantum and systematically controlled  
approach that captures the effects of weak disorder throughout the 
BCS-BEC crossover at leading order in the vicinity of $T_c$, without 
resorting to the local-density approximation or the replica trick.  
By expanding the effective action to second order in both the  
disorder potential and the bosonic field, and by retaining the  
necessary third- and fourth-order terms in the logarithmic expansion,  
our approach yields a controlled finite-temperature description  
that bridges the BCS and BEC regimes. In contrast to the $T = 0$  
approach, where disorder couples to bosonic fluctuations via the  
second-order term~\cite{orso07, iskin25c}, these contributions  
vanish near $T_c$, making higher-order terms essential.  
The formalism reproduces known limiting behaviors for weakly-interacting  
fermions and bosons, recovers the fermionic and bosonic self-energy  
diagrams of Ref.~\cite{palestini13} for the three-dimensional  
continuum model, and extends beyond them. Applicable to both continuum  
and lattice systems, the theory provides a natural route for  
describing the loss of phase coherence near $T_c$ and for connecting  
microscopic disorder models in ultracold gases with those relevant  
to superconductors.  

Experimental studies of disordered ultracold Fermi gases in the regime 
of static disorder across the BCS-BEC crossover remain relatively 
limited~\cite{nagler20,koch24}. 
Nevertheless, measurements of dipole oscillations in static 
speckle potentials have demonstrated that the damping coefficient 
depends sensitively on the interaction strength, indicating a 
nontrivial interplay between disorder and pairing 
correlations~\cite{nagler20}. 
More recently, experiments have explored the interaction-dependent 
stability of superfluid coherence under controlled optical speckle 
disorder, including studies of quenched-disorder dynamics~\cite{koch24}. 
These works highlight the strong sensitivity of interacting superfluids 
to disorder and demonstrate the feasibility of probing disorder effects 
across different interaction regimes. 
The present theory addresses the complementary regime of weak static 
disorder near $T_c$. In this limit, we predict a qualitative change in 
the disorder-induced shift of $T_c$ across the BCS-BEC crossover: a 
slight enhancement of $T_c$ on the BCS side and a suppression on the 
BEC side. This sign change provides a clear equilibrium benchmark for 
future experiments capable of measuring $T_c$ shifts under controlled 
weak disorder.

The paper is organized as follows. In Sec.~\ref{sec:eaa}, we derive the  
effective thermodynamic potential near $T_c$. In Sec.~\ref{sec:se}, we  
discuss the corresponding self-energy corrections for non-interacting  
Fermi and Bose gases. In Sec.~\ref{sec:ne}, we analyze the number equation  
and pair propagator in the BCS and BEC limits. In Sec.~\ref{sec:ct}, we  
examine disorder-induced corrections to $T_c$ across the crossover.  
Finally, in Sec.~\ref{sec:conc}, we summarize the results and outline future  
directions, with an Appendix detailing disorder corrections to the number  
equation in the limiting regimes.

\section{Effective-action approach near $T_c$}
\label{sec:eaa}

We assume that the disorder potential varies much more slowly than the
thermodynamic processes and it depends only on position.
For simplicity, we model it as a static, spin-independent white-noise 
potential and carry out the thermal average before performing the 
disorder average~\cite{cappellaro19, iskin25c}.
In momentum-frequency space, where $q = (\mathbf{q}, i q_\ell)$ 
with bosonic Matsubara frequencies
$q_\ell = 2\pi \ell / \beta$ for $\ell = 0, \pm 1, \pm 2, \cdots$, and
$\beta = 1/T$ is the inverse temperature in units of $k_\mathrm{B} \to 1$,
the disorder potential satisfies the correlations~\cite{orso07, iskin25c}
\begin{align}
\langle V_{q=0} \rangle_d &= 0, \\
\langle V_{q'} V_q \rangle_d &= \beta \kappa \delta_{q',-q} \delta_{q_\ell 0},
\end{align}
where $\langle \cdots \rangle_d$ denotes the statistical average over
random disorder realizations, $\kappa \ge 0$ quantifies the disorder strength,
and $\delta_{ij}$ is the Kronecker delta.

\subsection{Microscopic model}
\label{sec:mm}

Having a single-band continuum or a Hubbard model in mind, 
we consider the microscopic Hamiltonian~\cite{iskin25c}
\begin{align}
\mathcal{H} &= 
\sum_{\mathbf{k} \sigma} 
\xi_{\mathbf{k} \sigma}\,
c_{\mathbf{k} \sigma}^\dagger c_{\mathbf{k} \sigma}
+ \frac{1}{\sqrt{N}} 
\sum_{\mathbf{k} \mathbf{k'} \sigma} 
V_{\mathbf{k-k'}} \,
c_{\mathbf{k} \sigma}^\dagger c_{\mathbf{k'} \sigma}
\nonumber \\
&
- \frac{U}{N} 
\sum_{\mathbf{k k' q}} 
c_{\mathbf{k} + \mathbf{q}/2, \uparrow}^\dagger
c_{-\mathbf{k} + \mathbf{q}/2, \downarrow}^\dagger
c_{-\mathbf{k'} + \mathbf{q}/2, \downarrow}
c_{\mathbf{k'} + \mathbf{q}/2, \uparrow},
\end{align}
where $c_{\mathbf{k} \sigma}^\dagger$ ($c_{\mathbf{k} \sigma}$) 
creates (annihilates) a fermion with momentum $\mathbf{k}$ 
in units of $\hbar \to 1$ and spin $\sigma = \{\uparrow, \downarrow\}$. 
The shifted single-particle dispersion
$
\xi_{\mathbf{k} \sigma} = \varepsilon_{\mathbf{k} \sigma} - \mu
$
is measured relative to the chemical potential $\mu$, 
and satisfies both time-reversal symmetry
$
\varepsilon_{\mathbf{k}, \uparrow} = 
\varepsilon_{-\mathbf{k}, \downarrow} \equiv 
\varepsilon_{\mathbf{k}}
$
and inversion symmetry
$
\varepsilon_{\mathbf{k}} = \varepsilon_{-\mathbf{k}}.
$
We do not specify the explicit form of 
$
\xi_{\mathbf{k}} = \varepsilon_{\mathbf{k}} - \mu,
$
as our formulation applies generally to systems obeying the above symmetries. 
These conditions are satisfied, for example, in the continuum model with 
$\varepsilon_{\mathbf{k}} = |\mathbf{k}|^2 / (2m)$, 
in which case the number of lattice sites $N$ 
is replaced by the system volume $\mathcal{V}$. 
In the lattice model, we instead have
\begin{align}
\label{eqn:ek}
\varepsilon_\mathbf{k} = -\frac{1}{N}
\sum_{ij} t_{ij} e^{i \mathbf{k} \cdot \mathbf{r}_{ij}},
\end{align}
where $t_{ij}$ is the hopping parameter from site $j$ to site $i$
and 
$
\mathbf{r}_{ij} = \mathbf{r}_{j} - \mathbf{r}_{i}
$
denotes their relative position.
In the second term, $V_{\mathbf{q}} = V_q / \sqrt{\beta}$ 
denotes the Fourier component of the disorder potential. 
The attractive interaction between spin-$\uparrow$ and spin-$\downarrow$ 
fermions is taken to be contact-like (zero ranged) in the continuum model 
and onsite in the Hubbard model with coupling strength $U \ge 0$.

We next calculate the grand partition function $\mathcal{Z}$ 
for a given microscopic realization of the disorder potential, 
and subsequently perform the statistical average 
over all disorder realizations~\cite{orso07, cappellaro19, iskin25c}.

\subsection{Effective Gaussian action}
\label{sec:ga}

Using the functional-integral formalism in the momentum-frequency 
representation~\cite{sademelo93, engelbrecht97, taylor06, diener08, 
orso07, iskin25c}, the grand partition function can be expressed as
\begin{align}
\mathcal{Z} = \int \mathcal{D}[\phi, \phi^*] \, 
e^{-\mathcal{S}[\phi, \phi^*]},
\end{align}
where $\phi_q$ denotes the Fourier component of the complex bosonic 
field that captures fluctuations of the superfluid order parameter 
associated with Cooper pairing. 
In the vicinity of the critical temperature $T_c$ for the superfluid 
transition, the action can be written as~\cite{iskin25c}
\begin{align}
\mathcal{S}(\phi, \phi^*) = \frac{1}{U}\sum_q |\phi_q|^2 
- \mathrm{Tr}\ln(-\beta \mathbf{G}^{-1}),
\end{align}
which is a formally exact expression, where the trace $\mathrm{Tr}$ is 
taken over both Nambu and momentum-frequency spaces. 
Since the saddle-point order parameter vanishes 
(i.e., $\Delta \to 0$) near $T_c$, the inverse Nambu-Gor'kov Green’s 
function can be written as
$
\mathbf{G}^{-1}_{k k'} = \boldsymbol{\mathcal{G}}^{-1}_k \delta_{k k'} 
- \boldsymbol{\Sigma}_{k k'},
$
where 
\begin{align}
\boldsymbol{\mathcal{G}}_k &= 
\begin{pmatrix}
\mathcal{G}_k & 0 \\
0 & -\mathcal{G}_{-k} 
\end{pmatrix},
\\
\boldsymbol{\Sigma}_{k k'} &= 
\frac{1}{\sqrt{\beta N}}
\begin{pmatrix}
V_{k-k'} & -\phi_{k-k'} \\
-\phi_{k'-k}^* & -V_{k-k'}
\end{pmatrix}.
\end{align}
Here,
$
\mathcal{G}_k = 1/(ik_n - \xi_\mathbf{k})
$
is the non-interacting Green’s function, with
$
k \equiv (\mathbf{k}, i k_n),
$
and fermionic Matsubara frequencies $k_n = (2n+1)\pi/\beta$ for 
$n = 0, \pm 1, \pm 2,\cdots$. 

Noting that the non-interacting part is already diagonal in $k$ space, 
we can first rewrite
$
\mathbf{G}^{-1} = \boldsymbol{\mathcal{G}}^{-1} 
(\mathbf{1} - \boldsymbol{\mathcal{G}} \boldsymbol{\Sigma}),
$
where $\boldsymbol{\mathcal{G}}$ and $\boldsymbol{\Sigma}$ are
high-dimensional matrices in $k$ space, and then expand
$
\ln(\mathbf{1} - \boldsymbol{\mathcal{G}} \boldsymbol{\Sigma}) 
= - \sum_{n=1}^{\infty} \frac{(\boldsymbol{\mathcal{G}} \boldsymbol{\Sigma})^n}{n}
$
to the desired order.
In this work, we focus on the effects of weak disorder on 
the BCS-BEC crossover formalism. To this end, we must retain 
terms up to second order in both the disorder potential $V_q$ and 
the bosonic fields $\phi_q$. In contrast to the $T = 0$ case, where 
it suffices to consider the coupling between disorder and bosonic 
fluctuations originating from the second-order term in the 
logarithmic expansion~\cite{orso07, iskin25c}, these terms vanish 
near $T_c$. Therefore, the third- and fourth-order terms in the 
expansion must also be included as discussed below.

At the saddle-point level, the zeroth-order action in the logarithmic 
expansion
\begin{align}
\mathcal{S}_0 = - \sum_k \mathrm{tr}\ln(-\beta \boldsymbol{\mathcal{G}}^{-1}_k),
\end{align}
corresponds to the action of a non-interacting Fermi gas. 
The first-order action is purely a disorder term proportional 
to $V_{q=0}$ and therefore averages out from the effective thermodynamic 
potential. In the second-order action
\begin{align}
\mathcal{S}_2 = \frac{1}{2} \sum_{k_1 k_2} \mathrm{tr}
(\boldsymbol{\mathcal{G}}_{k_1} \boldsymbol{\Sigma}_{k_1 k_2}
\boldsymbol{\mathcal{G}}_{k_2} \boldsymbol{\Sigma}_{k_2 k_1}),
\end{align}
we separate the usual contribution from the disorder correction and write
$
\mathcal{S}_2 = \mathcal{S}_2^0 + \mathcal{S}_2^d.
$
The usual term can be expressed as
$
\mathcal{S}_2^0 = \sum_{q_1 q_2} A_{q_1 q_2} 
\phi_{q_1} \phi_{q_2}^*,
$
where 
$
A_{q_1 q_2} =  \Gamma_{q_1}^{-1} \delta_{q_1 q_2}
$
is diagonal in $q$ space, and 
$
\Gamma_q^{-1} = \frac{1}{U} - \frac{1}{\beta N} \sum_k
\mathcal{G}_{k+q} \mathcal{G}_{-k}
$
is the inverse fluctuation propagator in the absence of 
disorder~\cite{nsr85, sademelo93}. It can also be written as
\begin{align}
\label{eqn:Gammaq}
\Gamma_q^{-1} = \frac{1}{U} - \frac{1}{N} \sum_\mathbf{k}
\frac{1 - n_\mathrm{F}(\xi_\mathbf{k}) - n_\mathrm{F}(\xi_\mathbf{k+q})}
{iq_\ell - \xi_\mathbf{k} - \xi_\mathbf{k+q}},
\end{align}
where
$
n_\mathrm{F}(x) = 1/(e^{\beta x} + 1)
$
is the Fermi-Dirac distribution function. The saddle-point condition 
(i.e., the gap equation) near $T_c$ satisfies $\Gamma_0^{-1} = 0$, 
\begin{align}
\label{eqn:tc}
\frac{1}{U} = \frac{1}{N} \sum_\mathbf{k} 
\frac{1 - 2n_\mathrm{F}(\xi_\mathbf{k})}{2\xi_\mathbf{k}},
\end{align}
and remains unaffected by disorder~\cite{orso07, han11, iskin25c}.
Since the disorder correction to the second-order action,
$
\mathcal{S}_2^d = \frac{1}{2 \beta N} \sum_{k q} 
\mathrm{tr}\big(
\boldsymbol{\mathcal{G}}_k \boldsymbol{\tau}_z 
\boldsymbol{\mathcal{G}}_{k+q} \boldsymbol{\tau}_z
\big) V_{-q} V_q,
$
does not couple to $\phi_q$, the effective thermodynamic potential 
depends only on its disorder average,
$
\langle \mathcal{S}_2^d \rangle_d = \frac{\kappa}{2 N} \sum_{kq, q_\ell = 0}
(\mathcal{G}_k \mathcal{G}_{k+q} + \mathcal{G}_{-k} \mathcal{G}_{-k-q}).
$
Here $\boldsymbol{\tau}_z$ denotes the $z$ component of the Pauli 
matrices acting in the Nambu-Gor'kov space.
In the third-order action
\begin{align}
\mathcal{S}_3 = \frac{1}{3} \sum_{k_1 k_2 k_3} \mathrm{tr} 
(\boldsymbol{\mathcal{G}}_{k_1} \boldsymbol{\Sigma}_{k_1 k_2}
\boldsymbol{\mathcal{G}}_{k_2} \boldsymbol{\Sigma}_{k_2 k_3}
\boldsymbol{\mathcal{G}}_{k_3} \boldsymbol{\Sigma}_{k_3 k_1}),
\end{align}
keeping terms up to second order in $\phi_q$ and $V_q$, we find
$
\mathcal{S}_3^d = \sum_{q_1 q_2} B_{q_1 q_2} \phi_{q_1} \phi_{q_2}^*,
$
where
$
B_{q_1 q_2} = - \frac{2}{\sqrt{\beta N}^3} \sum_{k}
\mathcal{G}_{k+q_1} \mathcal{G}_{k+q_2} 
\mathcal{G}_{-k} 
V_{q_2-q_1}.
$
Similarly, in the fourth-order action
\begin{align}
\mathcal{S}_4 = \frac{1}{4} \sum_{k_1 k_2 k_3 k_4} \mathrm{tr} 
(\boldsymbol{\mathcal{G}}_{k_1} \boldsymbol{\Sigma}_{k_1 k_2}
\boldsymbol{\mathcal{G}}_{k_2} \boldsymbol{\Sigma}_{k_2 k_3}
\boldsymbol{\mathcal{G}}_{k_3} \boldsymbol{\Sigma}_{k_3 k_4}
\boldsymbol{\mathcal{G}}_{k_4} \boldsymbol{\Sigma}_{k_4 k_1}),
\end{align}
keeping terms up to second order in $\phi_q$ and $V_q$ through 
straightforward but lengthy algebra, we obtain
$
\mathcal{S}_4^d = \sum_{q_1 q_2}
C_{q_1 q_2} \phi_{q_1} \phi_{q_2}^*,
$
where
$
C_{q_1 q_2} = - \frac{1}{\beta^2 N^2} \sum_{k_1 k_2}
\big(
2\mathcal{G}_{k_2+q_2} \mathcal{G}_{k_1} 
\mathcal{G}_{k_2+q_1} \mathcal{G}_{-k_2}
+\mathcal{G}_{k_2+q_1} \mathcal{G}_{k_1} 
\mathcal{G}_{q_2-k_1} \mathcal{G}_{-k_2}
\big)
V_{k_2-k_1+q_2} V_{k_1-k_2-q_1}.
$
As a result, collecting all second-order bosonic terms together, 
we obtain
\begin{align}
\mathcal{S}_\mathrm{B} = \sum_{q_1 q_2} \Gamma_\mathrm{B}^{-1}(q_1,q_2) 
\phi_{q_1} \phi_{q_2}^*,
\end{align}
for the effective Gaussian action, where
$
\Gamma_\mathrm{B}^{-1}(q_1,q_2) = A_{q_1 q_2} + B_{q_1 q_2} + C_{q_1 q_2}
$
is the resulting inverse fluctuation propagator in the presence of weak disorder.

\subsection{Effective thermodynamic potential}
\label{sec:etp}

\begin{figure*} [htb]
\includegraphics[width = 0.7\textwidth]{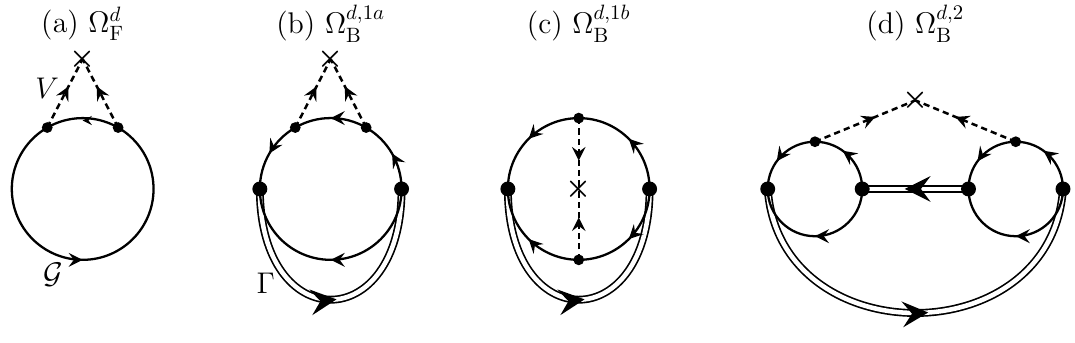}
\caption{\label{fig:FD} 
The diagrammatic representations of the disorder-induced thermodynamic 
potentials, given by Eqs.~(\ref{eqn:OFd}), (\ref{eqn:OBd1a}), 
(\ref{eqn:OBd1b}), and (\ref{eqn:OBd2}), offer an intuitive interpretation 
of the main theoretical results. In these diagrams, solid, dashed, and 
double lines represent $\mathcal{G}_k$, $V_q$, and $\Gamma_q$, respectively.
}
\end{figure*}

Next, we obtain the effective thermodynamic potential per lattice 
site~\cite{cappellaro19, orso07, iskin25c}
\begin{align}
\Omega_\mathrm{eff} = -\frac{\langle \ln \mathcal{Z}_\mathrm{eff} \rangle_d}
{\beta N},
\end{align}
by first integrating out the Gaussian bosonic degrees of freedom and then
averaging over the random disorder realizations. To make the discussion 
more tractable, we separate the fermionic and bosonic contributions and 
write
$
\Omega_\mathrm{eff} = \Omega_\mathrm{F} + \Omega_\mathrm{B}.
$
Furthermore, we separate the usual fermion term from its disorder 
correction and write
$
\Omega_\mathrm{F} = \Omega_\mathrm{F}^0 + \Omega_\mathrm{F}^d.
$ 
Here, the usual term 
$
\Omega_\mathrm{F}^0 = \mathcal{S}_0/(\beta N)
$ 
can be written as 
\begin{align}
\label{eqn:OF0}
\Omega_\mathrm{F}^0 &= \frac{2}{\beta N} \sum_\mathbf{k} 
\ln [n_\mathrm{F}(-\xi_\mathbf{k})], 
\end{align}
which corresponds to the thermodynamic potential of a non-interacting 
Fermi gas. Similarly, the disorder correction
$
\Omega_\mathrm{F}^d = \langle \mathcal{S}_2^d \rangle_d / (\beta N)
$
can be written as
\begin{align}
\label{eqn:OFd}
\Omega_\mathrm{F}^d &= \frac{\kappa}{N^2} 
\sum_{\mathbf{k}\mathbf{q}}
\frac{n_\mathrm{F}(\xi_\mathbf{k}) - n_\mathrm{F}(\xi_{\mathbf{k+q}})}
{\xi_\mathbf{k} - \xi_{\mathbf{k+q}}}.
\end{align}
Although its form appears quite different, the physical origin of this 
term is the same as that of the corresponding $T=0$ 
result~\cite{orso07, iskin25c}.
It corresponds to the Feynman diagram shown in Fig.~\ref{fig:FD}(a).

On the other hand, the bosonic contribution is given by
$
\Omega_\mathrm{B} = \frac{1}{\beta N} 
\langle \mathrm{Tr} \ln (\boldsymbol{\Gamma}_\mathrm{B}^{-1}/\beta) \rangle_d,
$
where we used the matrix identity
$
\ln \det \mathbf{M} = \mathrm{Tr} \ln \mathbf{M},
$
and $\boldsymbol{\Gamma}_\mathrm{B}^{-1}$ is a high-dimensional 
matrix representation in $q$ space. 
Noting that the usual term $\mathbf{A} \equiv \boldsymbol{\Gamma}^{-1}$ 
is already diagonal in $q$ space, we can first rewrite
$
\boldsymbol{\Gamma}_\mathrm{B}^{-1} = \boldsymbol{\Gamma}^{-1}
[\mathbf{1} + \boldsymbol{\Gamma} (\mathbf{B} + \mathbf{C})],
$
where $\mathbf{B}$ and $\mathbf{C}$ are again matrix representations 
in $q$ space, and then separate
$
\Omega_\mathrm{B} = \Omega_\mathrm{B}^0 + \Omega_\mathrm{B}^d.
$ 
Here, the zeroth-order term in the logarithmic expansion, 
\begin{align}
\label{eqn:OB0}
\Omega_\mathrm{B}^0 = - \frac{1}{\beta N} \sum_q 
\ln (\beta \Gamma_q),
\end{align}
corresponds to the usual bosonic contribution to the effective 
thermodynamic potential in the absence of 
disorder~\cite{nsr85, sademelo93}. 
Since our theory is accurate only up to second order in the disorder 
potential by construction, we expand the disorder correction
$
\Omega_\mathrm{B}^d = \frac{1}{\beta N} 
\langle \mathrm{Tr} \ln [\mathbf{1} + \boldsymbol{\Gamma} 
(\mathbf{B} + \mathbf{C})] \rangle_d
$
in powers of $\boldsymbol{\Gamma} (\mathbf{B} + \mathbf{C})$ 
and keep only the relevant terms. While the first-order term in 
$\boldsymbol{\Gamma} (\mathbf{B} + \mathbf{C})$ contributes as
$
\Omega_\mathrm{B}^{d,1} = \frac{1}{\beta N} 
\sum_q \langle \Gamma_q C_{qq} \rangle_d,
$
the second-order term contributes as
$
\Omega_\mathrm{B}^{d,2} = - \frac{1}{2 \beta N} 
\sum_{q k} \langle \Gamma_q B_{q,q+k} 
\Gamma_{q+k} B_{q+k,q} \rangle_d.
$
They can be written as
\begin{align}
\label{eqn:OBd1a}
\Omega_\mathrm{B}^{d,1a} &= -\frac{2\kappa}{\beta^2 N^3} 
\sum_{q k} \Gamma_q \mathcal{G}_{k}^2 \mathcal{G}_{q-k}
\sum_\mathbf{p} \mathcal{G}_{\mathbf{p-q}, k_n},
\\
\label{eqn:OBd1b}
\Omega_\mathrm{B}^{d,1b} &= -\frac{\kappa}{\beta^2 N^3} 
\sum_{q k} \Gamma_q \mathcal{G}_{q+k} \mathcal{G}_{-k}
\sum_\mathbf{p} \mathcal{G}_{\mathbf{q-p}, -k_n}
\mathcal{G}_{\mathbf{p}, q_\ell+k_n},
\\
\label{eqn:OBd2}
\Omega_\mathrm{B}^{d,2} &= -\frac{2\kappa}{\beta^3 N^4} 
\sum_{q k, k_n=0} \Gamma_q \Gamma_{q+k}
\bigg[\sum_p \mathcal{G}_{q+p} \mathcal{G}_{k+q+p} 
\mathcal{G}_{-p}\bigg]^2,
\end{align}
where 
$
\Omega_\mathrm{B}^{d,1} = \Omega_\mathrm{B}^{d,1a} + \Omega_\mathrm{B}^{d,1b}.
$
Unlike the analogous disorder correction at 
$T = 0$~\cite{orso07, iskin25c}, which has a remarkably simple 
form, these expressions are considerably more involved because 
they arise from a higher-order logarithmic expansion.
They correspond to the Feynman diagrams shown in 
Figs.~\ref{fig:FD}(b)-\ref{fig:FD}(d), which admit a clear and 
intuitive diagrammatic interpretation
~\footnote{
We are grateful to the anonymous referee for pointing out that 
each of our disorder contributions can be naturally associated 
with such intuitive Feynman diagrams
}.
In Sec.~\ref{sec:ne}, we show that while $\Omega_\mathrm{F}^d$ provides 
the leading-order disorder correction in the BCS limit, 
it is $\Omega_\mathrm{B}^{d,2}$ that dominates in the BEC limit.

\section{Self energy: BCS and BEC limits}
\label{sec:se}

For a non-interacting Fermi gas, the effective self-energy contribution 
that accounts for the effect of weak disorder at the lowest order in 
$\kappa$ is given by~\cite{abrikosov, palestini13}
\begin{align}
\Sigma_k^\mathrm{F} = \frac{\kappa}{N} \sum_\mathbf{p} 
\frac{1}{ik_n - \xi_{\mathbf{k+p}}}.
\end{align}
In particular, for the continuum model in three dimensions where 
we also set $N \to \mathcal{V}$, by first shifting 
$\mathbf{p} \to \mathbf{p-k}$ 
this expression can be rewritten as
$
\Sigma_k^\mathrm{F} 
= -m \kappa \Lambda/\pi^2 
+ m \kappa (ik_n + \mu) 
\int_0^\Lambda \frac{dp}{ik_n + \mu - \varepsilon_p},
$
where $\Lambda$ is the momentum cutoff. 
Since the latter integrand is symmetric, we rewrite it as half of 
an integral from $-\Lambda$ to $\Lambda$, take the limit 
$\Lambda \to \infty$, and express it as a contour integral,
$
\frac{1}{2} \oint \frac{dz}{ik_n + \mu - \varepsilon_z} 
= -i\pi \frac{\sqrt{m} \, \mathrm{sign}(k_n)}{\sqrt{2(ik_n + \mu)}},
$
where the contour is closed in the upper half-plane. 
When the imaginary part $\mathrm{Im} \sqrt{ik_n + \mu} \lessgtr 0$, 
i.e., for $k_n \lessgtr 0$, the poles $z_\mp = \mp \sqrt{2m(ik_n + \mu)}$ 
contribute with residues $\pm \sqrt{m/[2(ik_n + \mu)]}$, respectively. 
This leads to
\begin{align}
\label{eqn:seF}
\Sigma_k^\mathrm{F} 
= -\frac{m \kappa \Lambda}{\pi^2} 
- i\,\frac{\kappa (2m)^{3/2}}{4\pi} 
\sqrt{ik_n + \mu} \, \mathrm{sign}(k_n),
\end{align}
which is in perfect agreement with Eq.~(3) of Ref.~\cite{palestini13}. 
Noting that $|k_n| \ll \mu \ll \Lambda^2/(2m)$ in the BCS limit, 
the second term of $\Sigma_k^\mathrm{F}$ is small and can be 
neglected. The first term is referred to as $\Sigma_\mathbf{k}^\mathrm{F}$
in the rest of the paper. 
See Sec.~\ref{sec:neBCS} for further discussion.

Similarly, for a non-interacting Bose gas composed of tightly-bound 
fermion pairs, the effective self-energy contribution accounting for 
the lowest-order effect of disorder is given by~\cite{palestini13}
\begin{align}
Z\,\Sigma_q^M = \frac{\kappa_M}{N} \sum_\mathbf{p}
\frac{1}{iq_\ell - \xi_{\mathbf{q+p}}^M},
\end{align}
where $Z$ is the effective residue of the effective molecules,
$\kappa_M$ denotes their effective disorder strength, and
$\xi_\mathbf{q}^M = \varepsilon_\mathbf{q}^M - \mu_M$
is their effective dispersion, as discussed in Sec.~\ref{sec:pp}.
For the continuum model in three dimensions, we first follow
the same steps leading to Eq.~\eqref{eqn:seF}, and then note that
the real and imaginary parts of the square root satisfy
$
\mathrm{Re} \sqrt{iq_\ell + \mu_M}
= -\mathrm{sign}(q_\ell)\,
\mathrm{Im} \sqrt{-iq_\ell - \mu_M}
$
and
$
\mathrm{Im} \sqrt{iq_\ell + \mu_M}
= \mathrm{sign}(q_\ell)\,
\mathrm{Re} \sqrt{-iq_\ell - \mu_M}.
$
This yields
\begin{align}
\label{eqn:seB}
Z\,\Sigma_q^M
= -\frac{m_M \kappa_M \Lambda}{\pi^2}
+ \frac{\kappa_M (2m_M)^{3/2}}{4\pi}
\sqrt{-iq_\ell - \mu_M},
\end{align}
which is in perfect agreement with the point-boson result discussed 
in Ref.~\cite{lopatin02}. There is, however, a minor discrepancy with 
Eq.~(11) of Ref.~\cite{palestini13} due to an oversimplification in 
their second term.
In Secs.~\ref{sec:pp} and~\ref{sec:ct}, we argue that the diverging 
first term in Eq.~\eqref{eqn:seB} is absorbed into the Thouless
condition~\cite{lopatin02, palestini13},
$
\bar{\mu}_M = \mu_M + \frac{\kappa_M}{N}\sum_\mathbf{q}
\frac{1}{\varepsilon_\mathbf{q}^M} \to 0^+,
$
implying that the second term governs the disorder-induced
correction to $T_c$ in the BEC limit.
This stands in sharp contrast to the BCS limit,
where the diverging first term in Eq.~\eqref{eqn:seF} determines the 
corresponding correction to $T_c$, as discussed next.

\section{Number equation}
\label{sec:ne}

To ensure that our formalism remains applicable throughout the entire
BCS-BEC crossover, covering arbitrarily strong interaction strengths $U$,
we must solve Eq.~(\ref{eqn:tc}) self-consistently together with the
corresponding number 
equation~\cite{nsr85, sademelo93, engelbrecht97, taylor06, diener08,
orso07, iskin25c}.
In this work, we define the number of particles per lattice site as the
filling fraction $n = \mathcal{N}/N$, and obtain it through the 
thermodynamic relation
\begin{align}
\label{eqn:neff}
n = - \frac{\partial \Omega_\mathrm{eff}}{\partial \mu}\Big|_{\beta, N},
\end{align}
which ultimately leads to the number equation
$
n = n_\mathrm{F}^0 + n_\mathrm{F}^d + n_\mathrm{B}^0 + n_\mathrm{B}^d.
$

\subsection{Number equation in the BCS limit}
\label{sec:neBCS}

Since the bosonic fluctuations do not play a significant role in the
BCS limit, we expect
$
n \gg n_\mathrm{B}^0 + n_\mathrm{B}^d \to 0.
$
Furthermore, from Eq.~(\ref{eqn:OF0}), we find
\begin{align}
\label{eqn:nF0}
n_\mathrm{F}^0 = \frac{2}{N} \sum_\mathbf{k} n_\mathrm{F}(\xi_\mathbf{k}),
\end{align}
which corresponds to the standard number equation for a non-interacting
spin-$1/2$ Fermi gas. Similarly, from Eq.~(\ref{eqn:OFd}), we obtain
\begin{align}
\label{eqn:nFd}
n_\mathrm{F}^d = \frac{\kappa}{N^2} \sum_{\mathbf{k} \mathbf{q}}
\frac{n_\mathrm{F}'(\xi_\mathbf{k}) - n_\mathrm{F}'(\xi_\mathbf{k+q})}
{\xi_\mathbf{k} - \xi_\mathbf{k+q}},
\end{align}
which represents the lowest-order disorder correction to the number 
equation for a non-interacting Fermi gas, where
$
n_\mathrm{F}'(x) = \frac{d n_\mathrm{F}(x)}{dx}
= -\frac{\beta}{4} \mathrm{sech}^2(\beta x/2).
$

In the BCS limit, where
$
n \to n_\mathrm{F}^0 + n_\mathrm{F}^d,
$
the leading-order effects of weak disorder can be incorporated into 
the number equation through the self-energy as
\begin{align}
\label{eqn:nBCS}
n = \frac{2}{\beta N} \sum_k \frac{e^{i k_n 0^+}}
{ik_n - \xi_\mathbf{k} - \Sigma_k^\mathrm{F}}
\approx 2 \sum_\mathbf{k} n_\mathrm{F}(\xi_\mathbf{k} + \Sigma_\mathbf{k}^\mathrm{F}),
\end{align}
where we approximate $\Sigma_k^\mathrm{F}$ with its first term and 
refer to it as $\Sigma_\mathbf{k}^\mathrm{F}$ in the BCS limit, 
as already discussed in Sec.~\ref{sec:se}.
To make this connection explicit, we rewrite Eq.~(\ref{eqn:nFd}) as
$
n_\mathrm{F}^d = \frac{2}{N} \sum_\mathbf{k} n_\mathrm{F}'(\xi_\mathbf{k})
S_\mathbf{k},
$
where
$
S_\mathbf{k} = \frac{\kappa}{2N} \sum_\mathbf{q}
\big(
\frac{1}{\xi_\mathbf{k} - \xi_\mathbf{k+q}}
+ \frac{1}{\xi_\mathbf{k} - \xi_\mathbf{k-q}}
\big).
$
As an illustration, we now evaluate $S_\mathbf{k}$ in the BCS limit
for the continuum model in three dimensions.
Aligning $\mathbf{k}$ along the $z$ axis and averaging over the
directions of $\mathbf{q}$, we find
$
S_\mathbf{k} = -\frac{m}{2\pi^2 k} \int_0^{\Lambda} q
\ln\big| \frac{q+2k}{q-2k} \big| dq,
$
where $q = |\mathbf{q}|$ and $\Lambda \to \infty$ is the momentum cutoff.
Using the identity
$
\int_0^a x \ln\big| \frac{x+1}{x-1} \big| dx
= a + \frac{a^2-1}{2} \ln\big| \frac{a+1}{a-1} \big|
$
for $a > 1$, and taking the limit $\Lambda/(2k) \gg 1$, which is
appropriate for the BCS regime where the region $\xi_\mathbf{k} = 0$
provides the dominant contribution in $\mathbf{k}$ space, we obtain
$
S_\mathbf{k} = -m \kappa \Lambda / \pi^2.
$
This corresponds precisely to $\Sigma_\mathbf{k}^\mathrm{F}$ at the 
lowest order. See Appendix~\ref{sec:app} for further discussion.
In the theory of disordered metals, such a term is usually regarded
as an irrelevant constant that can be absorbed into a redefinition
of the chemical potential, i.e.,
$\bar{\mu} = \mu - \Sigma_\mathbf{k}^\mathrm{F}$.
However, within the context of the BCS-BEC crossover, this simplification
is no longer valid, since the renormalization of $\mu$ is an essential
part of the problem and must be treated explicitly~\cite{palestini13}.

It is important to clarify the scope of this analysis.
In Eq.~(\ref{eqn:nBCS}), the fermionic self-energy is
retained only through its real, momentum-independent contribution,
which yields an ultraviolet-divergent shift of the chemical potential.
The imaginary part of the self-energy, associated with quasiparticle
lifetime effects induced by disorder~\cite{palestini13}, is neglected 
within this leading-order near-$T_c$ treatment. While such lifetime effects
and the associated vertex corrections in the particle-particle ladder
(Cooperon) channel play a central role in microscopic diagrammatic
formulations of Anderson’s theorem for disordered BCS superconductors,
their self-consistent inclusion lies beyond the Gaussian fluctuation
framework employed here. Consequently, the present results should be
understood as describing the robustness of the number equation and the
transition temperature against weak disorder near $T_c$, rather than as
a full treatment of quasiparticle lifetimes or transport properties in
the weak-coupling BCS regime.

In contrast to the non-interacting Fermi gas, we next show that
$
n \to n_\mathrm{B}^0 + n_\mathrm{B}^d
\ne 2 \sum_\mathbf{q} n_\mathrm{B}(\xi_\mathbf{q}^M
+ Z \Sigma_\mathbf{q}^M)
$
for a non-interacting Bose gas composed of tightly-bound fermion pairs.
This inequality arises because the frequency-dependent part of 
the self-energy provides the dominant contribution, as already 
discussed in Sec.~\ref{sec:se}. Here,
$
n_\mathrm{B}(x) = 1/(e^{\beta x} - 1)
$
denotes the Bose-Einstein distribution function, and $\Sigma_\mathbf{q}^M$
refers to $\mathrm{Re}\,\Sigma_q^M$.

\subsection{Pair propagator in the BEC limit}
\label{sec:pp}

In the absence of disorder, when $\kappa = 0$, the inverse fluctuation 
propagator in Eq.~(\ref{eqn:Gammaq}) reduces to
$
\Gamma_q^{-1} = \frac{1}{U} + \frac{1}{N} \sum_\mathbf{k} 
\frac{1}{iq_\ell - \xi_\mathbf{k} - \xi_{\mathbf{k+q}}}
$
in the BEC limit, where $\mu \to -\varepsilon_b/2$~\cite{sademelo93}. 
Here, $\varepsilon_b \ge 0$ denotes the binding energy of the 
tightly-bound fermion pairs, determined from
\begin{align}
\label{eqn:Eb}
\frac{1}{U} = \frac{1}{N} \sum_\mathbf{k} 
\frac{1}{2\varepsilon_\mathbf{k} + \varepsilon_b}.
\end{align}
By expanding $\varepsilon_\mathbf{k+q}$ up to second order in $\mathbf{q}$, 
and using integration by parts
$
\sum_\mathbf{k} \frac{\partial^2 \varepsilon_\mathbf{k}}
{\partial k_i \partial k_j} \varepsilon_\mathbf{k} 
= - \sum_\mathbf{k} \frac{\partial \varepsilon_\mathbf{k}}{\partial k_i}
\frac{\partial \varepsilon_\mathbf{k}}{\partial k_j},
$
we obtain the effective pair propagator
\begin{align}
\label{eqn:Gammap}
\Gamma_q = \frac{Z}{iq_\ell - \xi_\mathbf{q}^M},
\end{align}
in the BEC limit, where
$
\xi_\mathbf{q}^M = \varepsilon_\mathbf{q}^M - \mu_M
$
is the effective dispersion for the fermion pairs and
$
\mu_M = 2\mu + \varepsilon_b
$
is their chemical potential. Here,
$
\varepsilon_\mathbf{q}^M = \frac{1}{2} \sum_{ij} q_i q_j (m_M^{-1})_{ij}
$
in the $\mathbf{q \to 0}$ limit, where
$
(m_M^{-1})_{ij} = \frac{2}{U N} \sum_\mathbf{k} 
\frac{\partial \varepsilon_\mathbf{k}}{\partial k_i} 
\frac{\partial \varepsilon_\mathbf{k}}{\partial k_j} 
$
is the effective inverse mass tensor for the lattice model~\cite{iskin25b} 
and
$
(m_M^{-1})_{ij} = \delta_{ij}/(2m)
$
is for the continuum model. At high $\mathbf{q}$, similar to 
Eq.~(\ref{eqn:ek}), we expect
$
\varepsilon_\mathbf{q}^M = \frac{1}{N}
\sum_{ij} t_{ij}^M (1 - e^{i \mathbf{q} \cdot \mathbf{r}_{ij}}) 
$
for the lattice model, where 
$t_{ij}^M = (m_M^{-1})_{ij}/(2a^2)$ is the
effective hopping parameter for the pairs from site $j$ to site $i$, 
and $a$ is the lattice spacing. 

In Eq.~(\ref{eqn:Gammap}), the effective residue of the tightly-bound 
fermion pairs can be determined via
$
\frac{1}{Z} = \frac{\partial \Gamma_q^{-1}}{\partial (iq_\ell)}\big|_{q = 0},
$
leading to 
$
\frac{1}{Z} = -\frac{1}{N} \sum_\mathbf{k} 
\frac{1}{(2\varepsilon_\mathbf{k} + \varepsilon_b)^2}.
$
For the continuum model in three dimensions, one finds 
$
Z = -8\pi\sqrt{\varepsilon_b/m^{3}}
$
with $\varepsilon_b = 1/(m a_s^2)$, where $a_s$ is the $s$-wave scattering 
length in vacuum~\cite{sademelo93, palestini13}, determined by
$
\frac{1}{U} = - \frac{m}{4\pi a_s} 
+ \frac{1}{N} \sum_\mathbf{k} \frac{1}{2\varepsilon_\mathbf{k}}.
$
For the lattice model, we find $Z = -\varepsilon_b^2$ with $\varepsilon_b = U$. 
Note that by plugging Eq.~(\ref{eqn:Eb}) into the expression 
$\Gamma_0^{-1} = 0$ in the BEC limit and taking the derivative of the 
resultant expression with respect to $\varepsilon_b$, we obtain
\begin{align}
\label{eqn:ZG}
\frac{1}{Z} = \frac{1}{\beta N} \sum_k 
\mathcal{G}_k^2 \mathcal{G}_{-k},
\end{align}
which is an alternative expression for the residue, and is particularly 
useful in the discussion of Sec.~\ref{sec:neBEC}.

Given the pair propagator Eq.~(\ref{eqn:Gammap}) for the clean 
BEC limit, the usual number of tightly-bound fermion pairs can 
be written as
$
n_{M}^0 = -\frac{1}{\beta N} \sum_q e^{i q_\ell 0^+} \frac{\Gamma_q}{Z},
$
in the absence of a disorder. 
The disorder induced self-energy expression Eq.~(\ref{eqn:seB}) 
can also be written as
$
Z \Sigma_q^M = \frac{\kappa_M}{N} \sum_{p, p_n = 0} \frac{\Gamma_{q+p}}{Z}.
$
Thus, the leading-order disorder correction to the number of tightly-bound 
fermion pairs is expected to be
$
n_{M}^d = -\frac{1}{\beta N} \sum_q 
e^{i q_\ell 0^+} \frac{\Gamma_q^2 \Sigma_q^M}{Z}.
$
In the presence of weak disorder, if we define 
$
\bar{\Gamma}_q^{-1} = \Gamma_q^{-1} - \Sigma_q^M,
$
which is equivalent to replacing
$
\mu_M \to \mu_M - Z \Sigma_q^M
$
in $\Gamma_q^{-1}$, we find 
$
n_{M} \to n_M^0 + n_M^d 
\approx -\frac{1}{\beta N} \sum_q e^{i q_\ell 0^+} \frac{\bar{\Gamma}_q}{Z}.
$
Note that the disorder-generalized Thouless condition 
$\bar{\Gamma}_0^{-1} = 0$ is also equivalent to $\mu_M = Z \Sigma_0^M$.
We next show how these expectations are met by our effective thermodynamic 
potential in the BEC limit.

\subsection{Number equation in the BEC limit}
\label{sec:neBEC}

Since the fermionic degrees of freedom play no significant role in the 
BEC limit, where the Fermi surface vanishes beyond a finite $U$,
as indicated by $\mu$ dropping below the band minimum toward 
$-\varepsilon_b/2$, we expect
$
n \gg n_\mathrm{F}^0 + n_\mathrm{F}^d \to 0.
$
Furthermore, from Eq.~(\ref{eqn:OB0}), we find
\begin{align}
\label{eqn:nB0}
n_\mathrm{B}^0 = \frac{2}{N} \sum_\mathbf{q} n_\mathrm{B}(\xi_\mathbf{q}^M),
\end{align}
which corresponds to twice the standard number equation for a 
non-interacting spin-$0$ Bose gas, i.e., $n_\mathrm{B}^0 = 2n_M^0$. 
Similarly, from Eqs.~(\ref{eqn:OBd1a}) -~(\ref{eqn:OBd2}), we
obtain the disorder corrections to the number equation in the 
BEC limit, noting that the dominant contribution originates from the
term containing the largest number of $\Gamma_q$ factors. 
For instance, in evaluating Eq.~(\ref{eqn:OBd2}) for
$\Omega_\mathrm{B}^{d,2}$, we assume that $|\mu| \gg \varepsilon_{\max}$,
where $\varepsilon_{\max}$ denotes the particle bandwidth, and that
$|\mu| \gg |q_\ell|$ over the relevant bosonic energy scales
~\footnote{
For the three-dimensional continuum model in the BEC limit, the
condition $|\mu| \gg \Lambda^2/(2m)$ guarantees that the 
characteristic pair size $\xi_{\mathrm{pair}}$ is much smaller 
than the disorder correlation length~\cite{palestini13}. Since
$\xi_{\mathrm{pair}} \propto a_s$ and $|\mu| \simeq 1/(2m a_s^2)$ 
in this regime, the required separation of length scales is 
automatically satisfied. An analogous argument applies to the 
lattice model, where in the strong-coupling limit $U/t \gg 1$ one
has $|\mu| \simeq U/2$ and $\varepsilon_{\max} = 12t$, while the 
pair size scales as $\xi_{\mathrm{pair}} \propto a\, t/U$. 
Under these conditions, the composite bosons respond to the disorder 
potential as effectively point-like particles
}.
Under these conditions, we may approximate
$
\sum_p \mathcal{G}_{q+p} \mathcal{G}_{k+q+p} \mathcal{G}_{-p} 
\to \sum_p \mathcal{G}_p^2 \mathcal{G}_{-p},
$
and use Eq.~(\ref{eqn:ZG}), which leads to
$
\Omega_\mathrm{B}^{d,2} \to -\frac{2\kappa}{\beta N^2} 
\sum_{q k, k_n=0} \frac{\Gamma_q \Gamma_{q+k}}{Z^2}.
$
Thus, we identify
\begin{align}
\label{eqn:nBd2}
n_\mathrm{B}^{d,2} = -\frac{2 \kappa_M}{\beta N^2}
\sum_{q k, k_n=0} \frac{\Gamma_q^2 \Gamma_{q+k}}{Z^3}
\end{align}
as the leading-order disorder correction of $\Omega_\mathrm{B}^{d,2}$ 
to the number equation, where $\kappa_M = 4\kappa$ and 
$
\partial (\Gamma_q/Z) / \partial \mu = -2 \Gamma_q^2/Z^2.
$
The factor of $4$ in $\kappa_M$ reflects that the 
effective disorder potential for a tightly-bound fermion pair 
is twice that of an unpaired fermion. 
Note that Eq.~(\ref{eqn:nBd2}) is 
equivalent to the self-energy correction
$
Z \Sigma_q^{\mathrm{B} d,2} = \frac{\kappa_M}{N} 
\sum_{k,k_n = 0} \frac{\Gamma_{q+k}}{Z},
$
which corresponds precisely to $\Sigma_{4i-4j}^B(q)$ in 
Ref.~\cite{palestini13}.

Similarly, in Eq.~(\ref{eqn:OBd1a}) for $\Omega_\mathrm{B}^{d,1a}$, 
we identify the sum over $\mathbf{p}$ as the BEC-limit form of the 
fermionic self-energy $\Sigma_k^\mathrm{F}$ in the $|\mu| \to \infty$
limit. To leading order, it can be approximated as
$
\frac{\kappa}{N} \sum_\mathbf{p} \mathcal{G}_{\mathbf{p-q}, k_n} 
\to \frac{\kappa \Lambda^3}{6\pi^2 \mu}
$
for the continuum model in three dimensions~\cite{palestini13}, 
leading to 
$
\Omega_\mathrm{B}^{d,1a} \to  
\frac{\kappa \Lambda^3}{3\pi^2 \beta N |\mu|} 
\sum_q \frac{\Gamma_q}{Z}.
$
For the lattice model, the same sum can instead be approximated as  
$
\frac{\kappa}{N} \sum_\mathbf{p} \mathcal{G}_{\mathbf{p-q}, k_n} 
\to \frac{\kappa}{\mu}
$
in the $U \to \infty$ limit. 
The continuum result agrees exactly with Eq.~(5) of Ref.~\cite{palestini13}. 
Hence, we obtain
\begin{align}
\label{eqn:nBd1a}
n_\mathrm{B}^{d,1a} = 
\frac{2 \kappa \Lambda^3}{3\pi^2 \beta N |\mu|}
\sum_q \frac{\Gamma_q^2}{Z^2},
\end{align}
as the leading-order disorder correction of $\Omega_\mathrm{B}^{d,1a}$ 
to the number equation for the continuum model. 
This expression is equivalent to the self-energy correction
$
Z \Sigma_q^{\mathrm{B} d,1a} = 
- \frac{\kappa \Lambda^3}{3\pi^2 |\mu|},
$
which corresponds to twice the $\Sigma_{4h}^B(q)$ term in 
Ref.~\cite{palestini13}. 
The factor of two arises because Ref.~\cite{palestini13} 
includes two such diagrams in their Fig.~4.

Thus, we conclude that while the leading-order self-energy diagrams 
associated with our bosonic contributions $\Omega_\mathrm{B}^{d,2}$ 
and $\Omega_\mathrm{B}^{d,1a}$ are properly included in  
Ref.~\cite{palestini13}, the diagram corresponding to the  
remaining bosonic term $\Omega_\mathrm{B}^{d,1b}$ is absent from  
their analysis. However, as discussed in  
Ref.~\cite{palestini13}, one finds that  
$
\Sigma_q^{\mathrm{B} d,1a} \ll \Sigma_q^{\mathrm{B} d,2}
$
in the BEC limit, and the same hierarchy holds for the remaining  
contribution
$
\Sigma_q^{\mathrm{B} d,1b}.
$
Therefore, the leading-order effects of weak disorder near $T_c$ can be
incorporated into the number equation through the self-energy as
\begin{align}
\label{eqn:nBEC}
n \to n_\mathrm{B}^0 + n_\mathrm{B}^d \approx 
-\frac{2}{\beta N} \sum_q \frac{e^{i q_\ell 0^+}}
{iq_\ell - \xi_\mathbf{q}^M - Z\Sigma_q^M},
\end{align}
where $\Sigma_q^M \to \Sigma_q^{\mathrm{B} d,2}$, or equivalently  
$n_\mathrm{B}^d \to n_\mathrm{B}^{d,2} = 2n_M^d$, in the BEC limit.  
See Appendix~\ref{sec:app} for further discussion.
As discussed in Sec.~\ref{sec:pp}, this corresponds to the expected  
number equation for a non-interacting Bose gas of tightly-bound  
fermion pairs, which can be used to reproduce the correct $T_c$,  
as shown next.

\section{Critical temperature $T_c$}
\label{sec:ct}

It turns out that the effect of the disorder potential on $T_c$ differs 
markedly between a weakly-interacting Fermi gas in the BCS limit and a 
weakly-interacting Bose gas of tightly-bound fermion pairs in the BEC 
limit~\cite{han11, palestini13}. 
While the self-energy expressions in Eqs.~(\ref{eqn:seF}) and~(\ref{eqn:seB}) 
may appear superficially similar, a key difference emerges: 
in the fermionic case, the chemical potential $\mu$ is large and positive 
at $T_c$, effectively shielding the system from disorder. 
As a result, $T_c$ is primarily determined by the saddle-point 
condition Eq.~(\ref{eqn:tc}), and Anderson’s theorem protects the 
transition temperature from weak disorder at the level of the saddle-point 
and Gaussian fluctuation analysis. In other words, the presence of 
a large Fermi surface restricts scattering processes, 
ensuring that weak disorder has only a minimal effect. 
In contrast, for tightly-bound fermion pairs in the BEC limit, the
effective bosonic chemical potential $\bar{\mu}_M \to 0^+$ at $T_c$, 
leaving them fully exposed to disorder. 
Since $T_c$ is essentially set by the number equation Eq.~(\ref{eqn:nBEC}), 
rather than by Eq.~(\ref{eqn:tc}) which fixes $\mu \to -\varepsilon_b/2$, 
weak disorder directly affects phase coherence through the emergence 
of an incoherent component in the pair propagator, characterized by the 
$\sqrt{-iq_\ell}$ dependence of the bosonic self-energy $\Sigma_q^M$. 
This distinction highlights the fundamental role of quantum statistics in 
determining the robustness of phase coherence against the disorder potential.

As an illustration, for the continuum model in three dimensions 
in the BCS limit, Eq.~(\ref{eqn:nBCS}) indicates that the chemical 
potential must be shifted as 
$
\mu \to \mu + \Sigma_\mathbf{k}^\mathrm{F}
$ 
to preserve the same particle filling $n$ as in the clean system. 
Because $\Sigma_\mathbf{k}^\mathrm{F} < 0$, this shift lowers $\mu$ 
relative to the clean case. Within the BCS-BEC crossover, a reduced 
chemical potential pushes the system toward the unitarity regime, 
where $T_c$ is higher. Therefore, inserting this lowered $\mu$ into 
Eq.~(\ref{eqn:tc}) naturally leads to a slightly enhanced $T_c$ 
compared to the clean system. Indeed, as reported by the numerical
calculations of Ref.~\cite{palestini13}, on the BCS side of unitarity, 
the presence of disorder produces an increase of $T_c$ by a few percent, 
indicating that weak disorder can favor fermion pairing in this regime.

On the other hand, in the BEC limit, direct evaluation of 
Eq.~(\ref{eqn:nB0}) gives
$
n_\mathrm{B}^0 = 2 [m_M T_c / (2\pi)]^{3/2} \zeta(3/2),
$
where $\zeta(3/2) = 2.612$ is the Riemann-zeta function, and the 
leading-order disorder correction is given by Eq.~(\ref{eqn:nBd2}). 
The latter can be calculated either by taking the derivative of 
Eq.~(\ref{eqn:seB}) with respect to $\mu_M$, or equivalently by 
performing the momentum sum over $\mathbf{q}$ via direct integration, 
using
$
\int_0^\infty \frac{\sqrt{\varepsilon}\, d\varepsilon}
{(iq_\ell - \varepsilon)^2} = \frac{\pi}{2\sqrt{-iq_\ell}}
$
for $q_\ell \ne 0$. This yields
$
n_\mathrm{B}^d = \frac{\kappa_M m_M^3}{2\pi^2} T_c,
$
where we set $\bar{\mu}_M \to 0$ and employ
$
\sum_{q_\ell \ne 0} e^{iq_\ell 0^+} = -1.
$
The latter identity follows from
$
\frac{1}{\beta} \sum_{q_\ell} e^{iq_\ell 0^+} 
= \frac{1}{2\pi i}\oint e^{z0^+} n_\mathrm{B}(z) dz = 0,
$
since the integrand vanishes along the circular periphery when the 
contour encircling the poles is extended to infinity. 
Note that the $q_\ell = 0$ frequency does not contribute to the 
disorder correction, and the arising $1/\sqrt{-iq_\ell}$ factor 
is exactly canceled by the $\sqrt{-iq_\ell}$ dependence of the 
bosonic self-energy. See Appendix~\ref{sec:app} for further discussion. 
As a result, we find
\begin{align}
\bigg(\frac{T_c}{T_{c,0}}\bigg)^{3/2} 
= 1 - \frac{\kappa_M m_M^3 T_c}{4\pi^2 n_{M,0}},
\end{align}
where
$
T_{c,0} = \frac{2\pi}{m_M} \big[\frac{n_{M,0}}{\zeta(3/2)}\big]^{2/3}
$
is the standard expression for the BEC transition temperature of a 
clean system, with $n_{M,0} = n/2$. To obtain the leading-order 
disorder correction, $T_c$ on the right-hand side can be 
replaced by $T_{c,0}$.
Thus, in sharp contrast to the slight enhancement of $T_c$ in the 
BCS limit, disorder leads to a pronounced suppression of $T_c$ in 
the BEC regime. This result is in excellent agreement with previous 
analyses of non-interacting point bosons at the lowest order in 
the disorder potential~\cite{lopatin02}, as well as with those 
for tightly-bound fermion pairs~\cite{han11, palestini13}.

\begin{figure}[htb]
\centerline{\scalebox{0.63}{\includegraphics{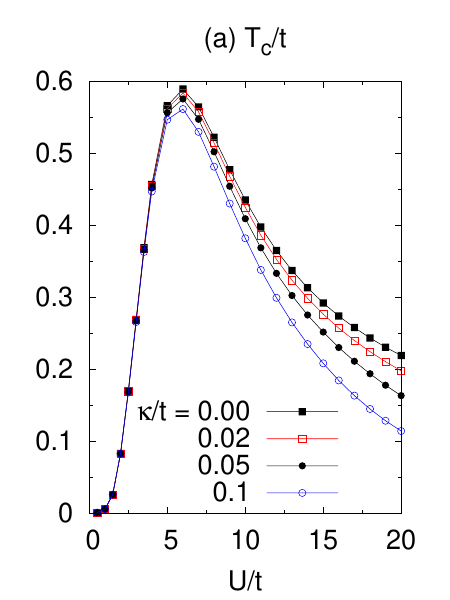} 
\hskip -0.8cm
\includegraphics{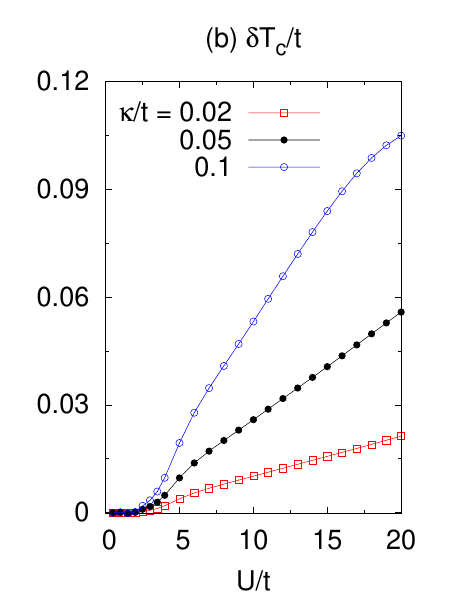}}}
\centerline{\scalebox{0.63}{\includegraphics{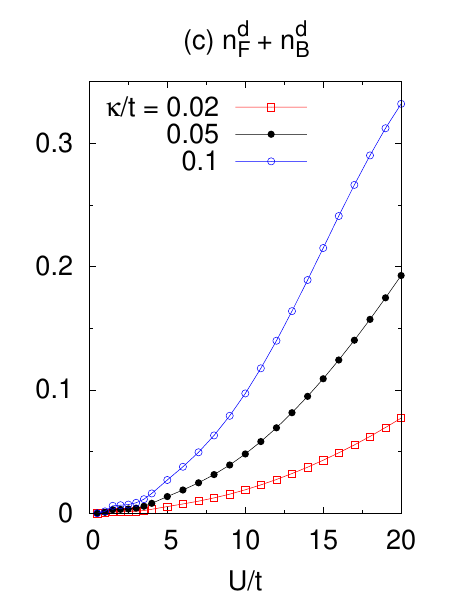}
\hskip -0.8cm
\includegraphics{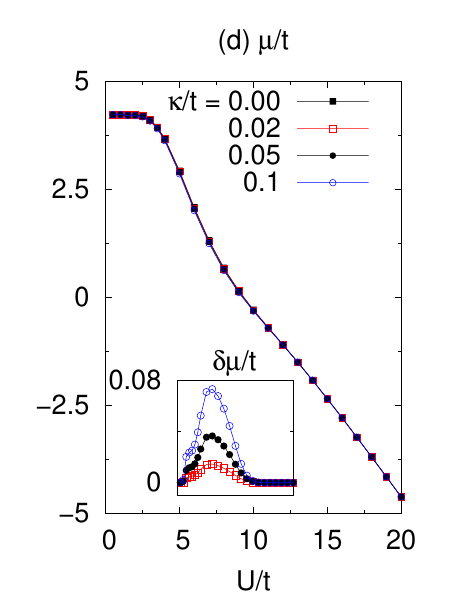}}}
\caption{\label{fig:Tc}
Self-consistent solutions for 
(a) the critical temperature $T_c$,
(b) its disorder suppression $\delta T_c = T_{c,0} - T_c$ 
with $T_{c,0}$ for the clean case,
(c) the disorder correction $n_\mathrm{F}^d + n_\mathrm{B}^d$ 
to the number equation, and 
(d) the chemical potential $\mu$.
The inset in (d) shows the corresponding disorder correction
$\delta \mu = \mu_0 - \mu$, with $\mu_0$ for the clean case.
All quantities are plotted as functions of the onsite interaction 
strength $U$ for $\kappa/t = \{0, 0.02, 0.05, 0.1\}$ at quarter 
filling ($n = 0.5$).
}
\end{figure}

In the BCS-BEC crossover problem, one must solve the saddle-point 
condition in Eq.~(\ref{eqn:tc}) together with the number equation 
in Eq.~(\ref{eqn:neff}) self-consistently in order to determine $T_c$ 
and $\mu$ as functions of $U$. Even in the clean case, this procedure 
is already nontrivial in the crossover regime~\cite{sademelo93}; 
the inclusion of disorder renders the problem substantially more 
involved due to the complicated structure of the corrections 
appearing in Eqs.~(\ref{eqn:OBd1a})-(\ref{eqn:OBd2}).
In Ref.~\cite{palestini13}, a comprehensive numerical analysis of 
the resulting leading-order self-energy diagrams was carried out 
across the BCS-BEC crossover under various approximations, together 
with analytic estimates in the extreme BCS and BEC limits. It was 
further shown that incorporating disorder effects at the level of 
Eqs.~(\ref{eqn:nBCS}) and~(\ref{eqn:nBEC}) provides a qualitatively 
reliable description in the intermediate regime~\cite{han11}, 
although quantitative predictions depend sensitively on the specific 
procedure used to treat cutoff-related issues~\cite{palestini13}.

As an illustration, in Fig.~\ref{fig:Tc} we consider a cubic lattice 
with dispersion
\begin{align}
\varepsilon_\mathbf{k} = 2t\left[3 - \cos(k_x a) 
- \cos(k_y a) - \cos(k_z a)\right],
\end{align}
where $a$ is the lattice spacing and $t > 0$ is the nearest-neighbor 
hopping amplitude. We present the self-consistent solutions for $T_c$, 
obtained by solving Eq.~(\ref{eqn:tc}) together with 
Eqs.~(\ref{eqn:nBCS}) and~(\ref{eqn:nBEC}) at quarter filling. 
Furthermore, we approximate the effective dispersion of fermion pairs as
$
\varepsilon_\mathbf{q}^M = \frac{4t^2}{U}\left[3 - \cos(q_x a) 
- \cos(q_y a) - \cos(q_z a)\right],
$
which corresponds to an effective nearest-neighbor hopping 
$t_M = 2t^2/U$ for the pairs (or, equivalently, an effective mass 
$m_M = U/(4t^2 a^2)$) and ensures that the interpolation scheme 
reproduces the correct BEC limit. Thus, while this numerical 
implementation captures both the BCS and BEC limits by construction, 
it provides only a qualitative description in the crossover regime. 
Consistent with our analysis above, the figure shows that the 
superfluid phase is most robust against disorder in the BCS 
and crossover regimes, whereas disorder significantly suppresses 
$T_c$ in the BEC regime. Although this trend is not clearly visible 
in the figures, we also observe a slight enhancement of $T_c$ in 
the weak-coupling BCS limit. For example, at $U = 0.5t$, we obtain 
$T_c \approx 0.000935t$ in the clean case, while 
$T_c/t \approx \{0.000946, 0.000950, 0.000956\}$ for 
$\kappa/t = \{0.02, 0.05, 0.1\}$, respectively. However, we note 
that our numerics do not converge with high precision at such small 
values of $U/t$. We further find that the chemical potential $\mu$ is 
reduced relative to its clean-limit value $\mu_0$ in the BCS regime, 
consistent with the discussion in Sec.~\ref{sec:neBCS}. In contrast, 
in the BEC regime $\mu$ is primarily determined by Eq.~(\ref{eqn:tc}), 
rather than by Eq.~(\ref{eqn:nBEC}), and therefore remains 
essentially unchanged from its clean-limit value.

Finally, we emphasize that while this Gaussian fluctuation framework 
provides a systematically controlled description near $T_c$ and 
recovers the correct BCS and BEC limits, it is not intended for a 
quantitatively exact treatment of the intermediate crossover. 
Since Gaussian theories often falter outside their controlled regimes, 
we employ this approximation to provide a consistent finite-temperature
interpolation anchored by analytical limits, rather than a full 
self-consistent description of spectral properties across the 
entire crossover.

\section{Conclusion}
\label{sec:conc}

In summary, we developed a systematic theoretical approach that 
incorporates the effects of weak disorder into the BCS-BEC crossover 
near the critical temperature $T_c$. Starting from a functional-integral 
formulation in momentum-frequency space, we derived an effective 
thermodynamic potential by expanding the action to second order in both 
the disorder potential and the bosonic field, which necessitates 
inclusion of third- and fourth-order terms in the logarithmic expansion 
near $T_c$. The resulting formalism fully captures Gaussian fluctuations 
of the order-parameter field and its coupling to a static white-noise 
disorder potential. It accurately reproduces the known BCS and 
BEC limits, recovering the fermionic and bosonic self-energy diagrams 
identified in Ref.~\cite{palestini13} for the three-dimensional 
continuum model, and extends beyond them. Within this controlled 
near-$T_c$ framework, the resulting thermodynamic potential and 
self-energy expressions provide a consistent leading-order description 
of the BCS-BEC crossover. By recovering established results for 
non-interacting point bosons and tightly-bound fermion pairs at the 
lowest order in disorder~\cite{lopatin02, palestini13}, this approach 
enables a systematic interpolation between the weak- and strong-coupling 
regimes.
The predicted crossover from disorder-enhanced to disorder-suppressed 
$T_c$ provides a clear benchmark for future cold-atom experiments with 
tunable speckle disorder. Systematic measurements of the transition 
temperature across the crossover could therefore directly test the 
self-consistent fluctuation-disorder interplay developed here.

The formalism applies to both continuum and lattice systems and can be  
naturally extended to multiband Hubbard models, providing a unified  
framework for studying the interplay between disorder and interactions.  
Such generalizations offer a promising route to investigate  
disorder-induced phenomena in multiband and quantum-geometric  
superfluids of strongly-correlated materials~\cite{torma22, yu24, jiang25}.  
The present results capture the leading-order corrections valid in the 
weak-disorder limit; however, exploring the regime of stronger disorder, 
where localization effects become significant, remains an important 
frontier for future investigation.

\begin{acknowledgments}
We thank G. C. Strinati for email correspondence. 
The author acknowledges funding from US Air Force Office of 
Scientific Research (AFOSR) Grant No. FA8655-24-1-7391.
\end{acknowledgments}

\appendix

\section{Disorder correction to the number equation 
in the BCS and BEC limits}
\label{sec:app}

For the continuum model in three dimensions, it is instructive to 
comment on the following observation. Since the bosonic self-energy 
$\Sigma_q^M$ depends on $q_\ell$ but not on $\mathbf{q}$, 
we may rewrite $n_M^d$ as 
$
\frac{1}{\beta N} \sum_{q_\ell} e^{i q_\ell 0^+} Z \Sigma_q^M
\sum_\mathbf{q} \frac{\partial}{\partial \mu_M} 
\big(\frac{1}{iq_\ell - \xi_\mathbf{q}^M}\big),
$
move the derivative outside the $\mathbf{q}$ sum, and use 
Eq.~(\ref{eqn:seB}) for the result of the $\mathbf{q}$ summation. 
After setting
$
\sum_{q_\ell \ne 0} e^{iq_\ell 0^+} = -1,
$
this procedure yields $\frac{\kappa_M m_M^3}{4 \pi^2} T_c$, 
in full agreement with the result obtained in Sec.~\ref{sec:ct}. 
Note that the $1/\sqrt{-iq_\ell}$ factor is canceled by the 
$\sqrt{-iq_\ell}$ dependence of $\Sigma_q^M$. 
Therefore, even though the $\mathbf{q}$ sum is formally divergent, 
the operations $\sum_\mathbf{q}$ and $\frac{\partial}{\partial \mu_M}$ 
happen to commute in this specific calculation.

In the BCS limit, a similar reasoning can be followed. 
Since the fermionic self-energy $\Sigma_k^\mathrm{F}$ depends on 
$k_n$ but not on $\mathbf{k}$, we may rewrite $n_\mathrm{F}^d$ as
$
-\frac{2}{\beta N} \sum_{k_n} e^{i k_n 0^+} \Sigma_k^\mathrm{F} 
\sum_\mathbf{k} \frac{\partial}{\partial \mu} 
\big( \frac{1}{ik_n - \xi_\mathbf{k}} \big),
$
move the derivative outside the $\mathbf{k}$ sum, and use the 
alternative expression
\begin{align}
\label{eqn:seFa}
\Sigma_k^\mathrm{F}
= -\frac{m \kappa \Lambda}{\pi^2}
+ \frac{\kappa (2m)^{3/2}}{4\pi}
\sqrt{-ik_n - \mu},
\end{align}
instead of Eq.~(\ref{eqn:seF}), for the $\mathbf{k}$ summation. 
After differentiation, this leads to
$
\frac{(2m)^{3/2}}{4\pi \beta} \sum_{k_n} 
\frac{e^{i k_n 0^+} \Sigma_k^\mathrm{F}}{\sqrt{-ik_n - \mu}}.
$
Substituting Eq.~(\ref{eqn:seFa}) produces two distinct terms. 
The second term vanishes because, when the $\sqrt{-ik_n - \mu}$ 
factors cancel, the frequency sum can be rewritten as a contour 
integral over $z = ik_n$, i.e., 
$
\frac{1}{\beta} \sum_{k_n} e^{i k_n 0^+} 
= -\frac{1}{2\pi i}\oint e^{z0^+} n_\mathrm{F}(z)\, dz,
$
which evaluates to zero since the integrand vanishes at the 
circular periphery when the contour enclosing the poles is 
extended to infinity. In contrast, for the first term, converting 
the sum into a contour integral, the integrand exhibits a branch 
cut for $\mathrm{Re}\, z > -\mu$, reducing the expression 
to an integral over the discontinuity 
$-2i/\sqrt{\varepsilon + \mu}$ across the cut:
\begin{align}
\sum_{k_n} 
\frac{e^{i k_n 0^+}}{\sqrt{-ik_n - \mu}} 
= \frac{-\beta}{2\pi i} \oint 
\frac{e^{z 0^+} n_\mathrm{F}(z)\, dz}{\sqrt{-z - \mu}}
=
\frac{\beta}{\pi} \int_{-\mu}^\infty  
\frac{n_\mathrm{F}(\varepsilon)\, d\varepsilon} 
{\sqrt{\varepsilon + \mu}}.
\end{align}
After integrating by parts, where the boundary terms vanish, the 
last expression becomes equivalent to the $\mathbf{k}$ sum
$
-\frac{8 \pi \beta}{(2m)^{3/2} N} 
\sum_\mathbf{k} n_\mathrm{F}'(\xi_\mathbf{k}).
$
This ultimately leads to
$
-\frac{2m\kappa \Lambda}{\pi^2 \beta N} 
\sum_\mathbf{k} \frac{\partial}{\partial \mu}
\sum_{k_n} \frac{e^{i k_n 0^+}}{ik_n - \xi_\mathbf{k}}
$
for $n_\mathrm{F}^d$, which differs by a minus sign from the correct 
result. This observation implies that, in this particular case, 
the operations $\sum_\mathbf{k}$ and $\frac{\partial}{\partial \mu}$ 
anticommute.

\bibliography{refs}

\begin{thebibliography}{35}%
\makeatletter
\providecommand \@ifxundefined [1]{%
 \@ifx{#1\undefined}
}%
\providecommand \@ifnum [1]{%
 \ifnum #1\expandafter \@firstoftwo
 \else \expandafter \@secondoftwo
 \fi
}%
\providecommand \@ifx [1]{%
 \ifx #1\expandafter \@firstoftwo
 \else \expandafter \@secondoftwo
 \fi
}%
\providecommand \natexlab [1]{#1}%
\providecommand \enquote  [1]{``#1''}%
\providecommand \bibnamefont  [1]{#1}%
\providecommand \bibfnamefont [1]{#1}%
\providecommand \citenamefont [1]{#1}%
\providecommand \href@noop [0]{\@secondoftwo}%
\providecommand \href [0]{\begingroup \@sanitize@url \@href}%
\providecommand \@href[1]{\@@startlink{#1}\@@href}%
\providecommand \@@href[1]{\endgroup#1\@@endlink}%
\providecommand \@sanitize@url [0]{\catcode `\\12\catcode `\$12\catcode `\&12\catcode `\#12\catcode `\^12\catcode `\_12\catcode `\%12\relax}%
\providecommand \@@startlink[1]{}%
\providecommand \@@endlink[0]{}%
\providecommand \url  [0]{\begingroup\@sanitize@url \@url }%
\providecommand \@url [1]{\endgroup\@href {#1}{\urlprefix }}%
\providecommand \urlprefix  [0]{URL }%
\providecommand \Eprint [0]{\href }%
\providecommand \doibase [0]{https://doi.org/}%
\providecommand \selectlanguage [0]{\@gobble}%
\providecommand \bibinfo  [0]{\@secondoftwo}%
\providecommand \bibfield  [0]{\@secondoftwo}%
\providecommand \translation [1]{[#1]}%
\providecommand \BibitemOpen [0]{}%
\providecommand \bibitemStop [0]{}%
\providecommand \bibitemNoStop [0]{.\EOS\space}%
\providecommand \EOS [0]{\spacefactor3000\relax}%
\providecommand \BibitemShut  [1]{\csname bibitem#1\endcsname}%
\let\auto@bib@innerbib\@empty
%</preamble>
\bibitem [{\citenamefont {Ospelkaus}\ \emph {et~al.}(2006)\citenamefont {Ospelkaus}, \citenamefont {Ospelkaus}, \citenamefont {Wille}, \citenamefont {Succo}, \citenamefont {Ernst}, \citenamefont {Sengstock},\ and\ \citenamefont {Bongs}}]{ospelkaus06}%
  \BibitemOpen
  \bibfield  {author} {\bibinfo {author} {\bibfnamefont {S.}~\bibnamefont {Ospelkaus}}, \bibinfo {author} {\bibfnamefont {C.}~\bibnamefont {Ospelkaus}}, \bibinfo {author} {\bibfnamefont {O.}~\bibnamefont {Wille}}, \bibinfo {author} {\bibfnamefont {M.}~\bibnamefont {Succo}}, \bibinfo {author} {\bibfnamefont {P.}~\bibnamefont {Ernst}}, \bibinfo {author} {\bibfnamefont {K.}~\bibnamefont {Sengstock}},\ and\ \bibinfo {author} {\bibfnamefont {K.}~\bibnamefont {Bongs}},\ }\bibfield  {title} {\bibinfo {title} {Localization of bosonic atoms by fermionic impurities in a three-dimensional optical lattice},\ }\href@noop {} {\bibfield  {journal} {\bibinfo  {journal} {Physical review letters}\ }\textbf {\bibinfo {volume} {96}},\ \bibinfo {pages} {180403} (\bibinfo {year} {2006})}\BibitemShut {NoStop}%
\bibitem [{\citenamefont {Modugno}(2010)}]{modugno10}%
  \BibitemOpen
  \bibfield  {author} {\bibinfo {author} {\bibfnamefont {G.}~\bibnamefont {Modugno}},\ }\bibfield  {title} {\bibinfo {title} {Anderson localization in {Bose--Einstein} condensates},\ }\href@noop {} {\bibfield  {journal} {\bibinfo  {journal} {Reports on progress in physics}\ }\textbf {\bibinfo {volume} {73}},\ \bibinfo {pages} {102401} (\bibinfo {year} {2010})}\BibitemShut {NoStop}%
\bibitem [{\citenamefont {Deissler}\ \emph {et~al.}(2010)\citenamefont {Deissler}, \citenamefont {Zaccanti}, \citenamefont {Roati}, \citenamefont {D’Errico}, \citenamefont {Fattori}, \citenamefont {Modugno}, \citenamefont {Modugno},\ and\ \citenamefont {Inguscio}}]{deissler10}%
  \BibitemOpen
  \bibfield  {author} {\bibinfo {author} {\bibfnamefont {B.}~\bibnamefont {Deissler}}, \bibinfo {author} {\bibfnamefont {M.}~\bibnamefont {Zaccanti}}, \bibinfo {author} {\bibfnamefont {G.}~\bibnamefont {Roati}}, \bibinfo {author} {\bibfnamefont {C.}~\bibnamefont {D’Errico}}, \bibinfo {author} {\bibfnamefont {M.}~\bibnamefont {Fattori}}, \bibinfo {author} {\bibfnamefont {M.}~\bibnamefont {Modugno}}, \bibinfo {author} {\bibfnamefont {G.}~\bibnamefont {Modugno}},\ and\ \bibinfo {author} {\bibfnamefont {M.}~\bibnamefont {Inguscio}},\ }\bibfield  {title} {\bibinfo {title} {Delocalization of a disordered bosonic system by repulsive interactions},\ }\href@noop {} {\bibfield  {journal} {\bibinfo  {journal} {Nature physics}\ }\textbf {\bibinfo {volume} {6}},\ \bibinfo {pages} {354} (\bibinfo {year} {2010})}\BibitemShut {NoStop}%
\bibitem [{\citenamefont {Sanchez-Palencia}\ and\ \citenamefont {Lewenstein}(2010)}]{sanchez10}%
  \BibitemOpen
  \bibfield  {author} {\bibinfo {author} {\bibfnamefont {L.}~\bibnamefont {Sanchez-Palencia}}\ and\ \bibinfo {author} {\bibfnamefont {M.}~\bibnamefont {Lewenstein}},\ }\bibfield  {title} {\bibinfo {title} {Disordered quantum gases under control},\ }\href@noop {} {\bibfield  {journal} {\bibinfo  {journal} {Nature Physics}\ }\textbf {\bibinfo {volume} {6}},\ \bibinfo {pages} {87} (\bibinfo {year} {2010})}\BibitemShut {NoStop}%
\bibitem [{\citenamefont {Kondov}\ \emph {et~al.}(2011)\citenamefont {Kondov}, \citenamefont {McGehee}, \citenamefont {Zirbel},\ and\ \citenamefont {DeMarco}}]{kondov11}%
  \BibitemOpen
  \bibfield  {author} {\bibinfo {author} {\bibfnamefont {S.}~\bibnamefont {Kondov}}, \bibinfo {author} {\bibfnamefont {W.}~\bibnamefont {McGehee}}, \bibinfo {author} {\bibfnamefont {J.}~\bibnamefont {Zirbel}},\ and\ \bibinfo {author} {\bibfnamefont {B.}~\bibnamefont {DeMarco}},\ }\bibfield  {title} {\bibinfo {title} {Three-dimensional {Anderson} localization of ultracold matter},\ }\href@noop {} {\bibfield  {journal} {\bibinfo  {journal} {Science}\ }\textbf {\bibinfo {volume} {334}},\ \bibinfo {pages} {66} (\bibinfo {year} {2011})}\BibitemShut {NoStop}%
\bibitem [{\citenamefont {Jendrzejewski}\ \emph {et~al.}(2012)\citenamefont {Jendrzejewski}, \citenamefont {Bernard}, \citenamefont {Mueller}, \citenamefont {Cheinet}, \citenamefont {Josse}, \citenamefont {Piraud}, \citenamefont {Pezz{\'e}}, \citenamefont {Sanchez-Palencia}, \citenamefont {Aspect},\ and\ \citenamefont {Bouyer}}]{jendrzejewski12}%
  \BibitemOpen
  \bibfield  {author} {\bibinfo {author} {\bibfnamefont {F.}~\bibnamefont {Jendrzejewski}}, \bibinfo {author} {\bibfnamefont {A.}~\bibnamefont {Bernard}}, \bibinfo {author} {\bibfnamefont {K.}~\bibnamefont {Mueller}}, \bibinfo {author} {\bibfnamefont {P.}~\bibnamefont {Cheinet}}, \bibinfo {author} {\bibfnamefont {V.}~\bibnamefont {Josse}}, \bibinfo {author} {\bibfnamefont {M.}~\bibnamefont {Piraud}}, \bibinfo {author} {\bibfnamefont {L.}~\bibnamefont {Pezz{\'e}}}, \bibinfo {author} {\bibfnamefont {L.}~\bibnamefont {Sanchez-Palencia}}, \bibinfo {author} {\bibfnamefont {A.}~\bibnamefont {Aspect}},\ and\ \bibinfo {author} {\bibfnamefont {P.}~\bibnamefont {Bouyer}},\ }\bibfield  {title} {\bibinfo {title} {Three-dimensional localization of ultracold atoms in an optical disordered potential},\ }\href@noop {} {\bibfield  {journal} {\bibinfo  {journal} {Nature Physics}\ }\textbf {\bibinfo {volume} {8}},\ \bibinfo {pages} {398} (\bibinfo {year} {2012})}\BibitemShut {NoStop}%
\bibitem [{\citenamefont {Krinner}\ \emph {et~al.}(2013)\citenamefont {Krinner}, \citenamefont {Stadler}, \citenamefont {Meineke}, \citenamefont {Brantut},\ and\ \citenamefont {Esslinger}}]{krinner13}%
  \BibitemOpen
  \bibfield  {author} {\bibinfo {author} {\bibfnamefont {S.}~\bibnamefont {Krinner}}, \bibinfo {author} {\bibfnamefont {D.}~\bibnamefont {Stadler}}, \bibinfo {author} {\bibfnamefont {J.}~\bibnamefont {Meineke}}, \bibinfo {author} {\bibfnamefont {J.-P.}\ \bibnamefont {Brantut}},\ and\ \bibinfo {author} {\bibfnamefont {T.}~\bibnamefont {Esslinger}},\ }\bibfield  {title} {\bibinfo {title} {Superfluidity with disorder in a thin film of quantum gas},\ }\href {https://doi.org/10.1103/PhysRevLett.110.100601} {\bibfield  {journal} {\bibinfo  {journal} {Phys. Rev. Lett.}\ }\textbf {\bibinfo {volume} {110}},\ \bibinfo {pages} {100601} (\bibinfo {year} {2013})}\BibitemShut {NoStop}%
\bibitem [{\citenamefont {Krinner}\ \emph {et~al.}(2015)\citenamefont {Krinner}, \citenamefont {Stadler}, \citenamefont {Meineke}, \citenamefont {Brantut},\ and\ \citenamefont {Esslinger}}]{krinner15}%
  \BibitemOpen
  \bibfield  {author} {\bibinfo {author} {\bibfnamefont {S.}~\bibnamefont {Krinner}}, \bibinfo {author} {\bibfnamefont {D.}~\bibnamefont {Stadler}}, \bibinfo {author} {\bibfnamefont {J.}~\bibnamefont {Meineke}}, \bibinfo {author} {\bibfnamefont {J.-P.}\ \bibnamefont {Brantut}},\ and\ \bibinfo {author} {\bibfnamefont {T.}~\bibnamefont {Esslinger}},\ }\bibfield  {title} {\bibinfo {title} {Observation of a fragmented, strongly interacting {Fermi} gas},\ }\href {https://doi.org/10.1103/PhysRevLett.115.045302} {\bibfield  {journal} {\bibinfo  {journal} {Phys. Rev. Lett.}\ }\textbf {\bibinfo {volume} {115}},\ \bibinfo {pages} {045302} (\bibinfo {year} {2015})}\BibitemShut {NoStop}%
\bibitem [{\citenamefont {Choi}\ \emph {et~al.}(2016)\citenamefont {Choi}, \citenamefont {Hild}, \citenamefont {Zeiher}, \citenamefont {Schau{\ss}}, \citenamefont {Rubio-Abadal}, \citenamefont {Yefsah}, \citenamefont {Khemani}, \citenamefont {Huse}, \citenamefont {Bloch},\ and\ \citenamefont {Gross}}]{choi16}%
  \BibitemOpen
  \bibfield  {author} {\bibinfo {author} {\bibfnamefont {J.-y.}\ \bibnamefont {Choi}}, \bibinfo {author} {\bibfnamefont {S.}~\bibnamefont {Hild}}, \bibinfo {author} {\bibfnamefont {J.}~\bibnamefont {Zeiher}}, \bibinfo {author} {\bibfnamefont {P.}~\bibnamefont {Schau{\ss}}}, \bibinfo {author} {\bibfnamefont {A.}~\bibnamefont {Rubio-Abadal}}, \bibinfo {author} {\bibfnamefont {T.}~\bibnamefont {Yefsah}}, \bibinfo {author} {\bibfnamefont {V.}~\bibnamefont {Khemani}}, \bibinfo {author} {\bibfnamefont {D.~A.}\ \bibnamefont {Huse}}, \bibinfo {author} {\bibfnamefont {I.}~\bibnamefont {Bloch}},\ and\ \bibinfo {author} {\bibfnamefont {C.}~\bibnamefont {Gross}},\ }\bibfield  {title} {\bibinfo {title} {Exploring the many-body localization transition in two dimensions},\ }\href@noop {} {\bibfield  {journal} {\bibinfo  {journal} {Science}\ }\textbf {\bibinfo {volume} {352}},\ \bibinfo {pages} {1547} (\bibinfo {year} {2016})}\BibitemShut {NoStop}%
\bibitem [{\citenamefont {Cappellaro}\ and\ \citenamefont {Salasnich}(2019)}]{cappellaro19}%
  \BibitemOpen
  \bibfield  {author} {\bibinfo {author} {\bibfnamefont {A.}~\bibnamefont {Cappellaro}}\ and\ \bibinfo {author} {\bibfnamefont {L.}~\bibnamefont {Salasnich}},\ }\bibfield  {title} {\bibinfo {title} {Superfluids, fluctuations and disorder},\ }\href@noop {} {\bibfield  {journal} {\bibinfo  {journal} {Applied Sciences}\ }\textbf {\bibinfo {volume} {9}},\ \bibinfo {pages} {1498} (\bibinfo {year} {2019})}\BibitemShut {NoStop}%
\bibitem [{\citenamefont {Nagler}\ \emph {et~al.}(2020)\citenamefont {Nagler}, \citenamefont {J\"agering}, \citenamefont {Sheikhan}, \citenamefont {Barbosa}, \citenamefont {Koch}, \citenamefont {Eggert}, \citenamefont {Schneider},\ and\ \citenamefont {Widera}}]{nagler20}%
  \BibitemOpen
  \bibfield  {author} {\bibinfo {author} {\bibfnamefont {B.}~\bibnamefont {Nagler}}, \bibinfo {author} {\bibfnamefont {K.}~\bibnamefont {J\"agering}}, \bibinfo {author} {\bibfnamefont {A.}~\bibnamefont {Sheikhan}}, \bibinfo {author} {\bibfnamefont {S.}~\bibnamefont {Barbosa}}, \bibinfo {author} {\bibfnamefont {J.}~\bibnamefont {Koch}}, \bibinfo {author} {\bibfnamefont {S.}~\bibnamefont {Eggert}}, \bibinfo {author} {\bibfnamefont {I.}~\bibnamefont {Schneider}},\ and\ \bibinfo {author} {\bibfnamefont {A.}~\bibnamefont {Widera}},\ }\bibfield  {title} {\bibinfo {title} {Dipole oscillations of fermionic quantum gases along the {BEC-BCS} crossover in disordered potentials},\ }\href {https://doi.org/10.1103/PhysRevA.101.053633} {\bibfield  {journal} {\bibinfo  {journal} {Phys. Rev. A}\ }\textbf {\bibinfo {volume} {101}},\ \bibinfo {pages} {053633} (\bibinfo {year} {2020})}\BibitemShut {NoStop}%
\bibitem [{\citenamefont {Koch}\ \emph {et~al.}(2024)\citenamefont {Koch}, \citenamefont {Barbosa}, \citenamefont {Lang},\ and\ \citenamefont {Widera}}]{koch24}%
  \BibitemOpen
  \bibfield  {author} {\bibinfo {author} {\bibfnamefont {J.}~\bibnamefont {Koch}}, \bibinfo {author} {\bibfnamefont {S.}~\bibnamefont {Barbosa}}, \bibinfo {author} {\bibfnamefont {F.}~\bibnamefont {Lang}},\ and\ \bibinfo {author} {\bibfnamefont {A.}~\bibnamefont {Widera}},\ }\bibfield  {title} {\bibinfo {title} {Stability and sensitivity of interacting fermionic superfluids to quenched disorder},\ }\href@noop {} {\bibfield  {journal} {\bibinfo  {journal} {Nature Communications}\ }\textbf {\bibinfo {volume} {15}},\ \bibinfo {pages} {9292} (\bibinfo {year} {2024})}\BibitemShut {NoStop}%
\bibitem [{\citenamefont {Russ}\ \emph {et~al.}(2025)\citenamefont {Russ}, \citenamefont {Yan}, \citenamefont {Kowalski}, \citenamefont {Wadleigh}, \citenamefont {Scarola},\ and\ \citenamefont {DeMarco}}]{russ25}%
  \BibitemOpen
  \bibfield  {author} {\bibinfo {author} {\bibfnamefont {P.}~\bibnamefont {Russ}}, \bibinfo {author} {\bibfnamefont {M.}~\bibnamefont {Yan}}, \bibinfo {author} {\bibfnamefont {N.}~\bibnamefont {Kowalski}}, \bibinfo {author} {\bibfnamefont {L.}~\bibnamefont {Wadleigh}}, \bibinfo {author} {\bibfnamefont {V.~W.}\ \bibnamefont {Scarola}},\ and\ \bibinfo {author} {\bibfnamefont {B.}~\bibnamefont {DeMarco}},\ }\href {https://arxiv.org/abs/2506.16466} {\bibinfo {title} {Compressibility measurement of the thermal {MI--BG} transition in an optical lattice}} (\bibinfo {year} {2025}),\ \Eprint {https://arxiv.org/abs/2506.16466} {arXiv:2506.16466 [cond-mat.quant-gas]} \BibitemShut {NoStop}%
\bibitem [{\citenamefont {Giorgini}\ \emph {et~al.}(2008)\citenamefont {Giorgini}, \citenamefont {Pitaevskii},\ and\ \citenamefont {Stringari}}]{giorgini08}%
  \BibitemOpen
  \bibfield  {author} {\bibinfo {author} {\bibfnamefont {S.}~\bibnamefont {Giorgini}}, \bibinfo {author} {\bibfnamefont {L.~P.}\ \bibnamefont {Pitaevskii}},\ and\ \bibinfo {author} {\bibfnamefont {S.}~\bibnamefont {Stringari}},\ }\bibfield  {title} {\bibinfo {title} {Theory of ultracold atomic {Fermi} gases},\ }\href {https://doi.org/10.1103/RevModPhys.80.1215} {\bibfield  {journal} {\bibinfo  {journal} {Rev. Mod. Phys.}\ }\textbf {\bibinfo {volume} {80}},\ \bibinfo {pages} {1215} (\bibinfo {year} {2008})}\BibitemShut {NoStop}%
\bibitem [{\citenamefont {Ketterle}\ and\ \citenamefont {Zwierlein}(2008)}]{ketterle08}%
  \BibitemOpen
  \bibfield  {author} {\bibinfo {author} {\bibfnamefont {W.}~\bibnamefont {Ketterle}}\ and\ \bibinfo {author} {\bibfnamefont {M.~W.}\ \bibnamefont {Zwierlein}},\ }\bibfield  {title} {\bibinfo {title} {Making, probing and understanding ultracold {Fermi} gases},\ }\href@noop {} {\bibfield  {journal} {\bibinfo  {journal} {La Rivista del Nuovo Cimento}\ }\textbf {\bibinfo {volume} {31}},\ \bibinfo {pages} {247} (\bibinfo {year} {2008})}\BibitemShut {NoStop}%
\bibitem [{\citenamefont {Strinati}\ \emph {et~al.}(2018)\citenamefont {Strinati}, \citenamefont {Pieri}, \citenamefont {R{\"o}pke}, \citenamefont {Schuck},\ and\ \citenamefont {Urban}}]{strinati18}%
  \BibitemOpen
  \bibfield  {author} {\bibinfo {author} {\bibfnamefont {G.~C.}\ \bibnamefont {Strinati}}, \bibinfo {author} {\bibfnamefont {P.}~\bibnamefont {Pieri}}, \bibinfo {author} {\bibfnamefont {G.}~\bibnamefont {R{\"o}pke}}, \bibinfo {author} {\bibfnamefont {P.}~\bibnamefont {Schuck}},\ and\ \bibinfo {author} {\bibfnamefont {M.}~\bibnamefont {Urban}},\ }\bibfield  {title} {\bibinfo {title} {The {BCS--BEC} crossover: From ultra-cold {Fermi} gases to nuclear systems},\ }\href@noop {} {\bibfield  {journal} {\bibinfo  {journal} {Physics Reports}\ }\textbf {\bibinfo {volume} {738}},\ \bibinfo {pages} {1} (\bibinfo {year} {2018})}\BibitemShut {NoStop}%
\bibitem [{\citenamefont {Hartke}\ \emph {et~al.}(2025)\citenamefont {Hartke}, \citenamefont {Oreg}, \citenamefont {Feng}, \citenamefont {Turnbaugh}, \citenamefont {Hertkorn}, \citenamefont {He}, \citenamefont {Jia}, \citenamefont {Khatami}, \citenamefont {Zhang},\ and\ \citenamefont {Zwierlein}}]{hartke25}%
  \BibitemOpen
  \bibfield  {author} {\bibinfo {author} {\bibfnamefont {T.}~\bibnamefont {Hartke}}, \bibinfo {author} {\bibfnamefont {B.}~\bibnamefont {Oreg}}, \bibinfo {author} {\bibfnamefont {C.}~\bibnamefont {Feng}}, \bibinfo {author} {\bibfnamefont {C.}~\bibnamefont {Turnbaugh}}, \bibinfo {author} {\bibfnamefont {J.}~\bibnamefont {Hertkorn}}, \bibinfo {author} {\bibfnamefont {Y.-Y.}\ \bibnamefont {He}}, \bibinfo {author} {\bibfnamefont {N.}~\bibnamefont {Jia}}, \bibinfo {author} {\bibfnamefont {E.}~\bibnamefont {Khatami}}, \bibinfo {author} {\bibfnamefont {S.}~\bibnamefont {Zhang}},\ and\ \bibinfo {author} {\bibfnamefont {M.}~\bibnamefont {Zwierlein}},\ }\href {https://arxiv.org/abs/2511.10605} {\bibinfo {title} {Competition of fermion pairing, magnetism, and charge order in the spin-doped attractive {Hubbard} gas}} (\bibinfo {year} {2025}),\ \Eprint {https://arxiv.org/abs/2511.10605} {arXiv:2511.10605 [cond-mat.quant-gas]} \BibitemShut {NoStop}%
\bibitem [{\citenamefont {Orso}(2007)}]{orso07}%
  \BibitemOpen
  \bibfield  {author} {\bibinfo {author} {\bibfnamefont {G.}~\bibnamefont {Orso}},\ }\bibfield  {title} {\bibinfo {title} {{BCS-BEC} crossover in a random external potential},\ }\href {https://doi.org/10.1103/PhysRevLett.99.250402} {\bibfield  {journal} {\bibinfo  {journal} {Phys. Rev. Lett.}\ }\textbf {\bibinfo {volume} {99}},\ \bibinfo {pages} {250402} (\bibinfo {year} {2007})}\BibitemShut {NoStop}%
\bibitem [{\citenamefont {Han}\ and\ \citenamefont {S\'a~de Melo}(2011)}]{han11}%
  \BibitemOpen
  \bibfield  {author} {\bibinfo {author} {\bibfnamefont {L.}~\bibnamefont {Han}}\ and\ \bibinfo {author} {\bibfnamefont {C.~A.~R.}\ \bibnamefont {S\'a~de Melo}},\ }\bibfield  {title} {\bibinfo {title} {Evolution from {Bardeen--Cooper--Schrieffer} to {Bose--Einstein} condensate superfluidity in the presence of disorder},\ }\href@noop {} {\bibfield  {journal} {\bibinfo  {journal} {New Journal of Physics}\ }\textbf {\bibinfo {volume} {13}},\ \bibinfo {pages} {055012} (\bibinfo {year} {2011})}\BibitemShut {NoStop}%
\bibitem [{\citenamefont {Palestini}\ and\ \citenamefont {Strinati}(2013)}]{palestini13}%
  \BibitemOpen
  \bibfield  {author} {\bibinfo {author} {\bibfnamefont {F.}~\bibnamefont {Palestini}}\ and\ \bibinfo {author} {\bibfnamefont {G.~C.}\ \bibnamefont {Strinati}},\ }\bibfield  {title} {\bibinfo {title} {Systematic investigation of the effects of disorder at the lowest order throughout the {BCS-BEC} crossover},\ }\href {https://doi.org/10.1103/PhysRevB.88.174504} {\bibfield  {journal} {\bibinfo  {journal} {Phys. Rev. B}\ }\textbf {\bibinfo {volume} {88}},\ \bibinfo {pages} {174504} (\bibinfo {year} {2013})}\BibitemShut {NoStop}%
\bibitem [{\citenamefont {Iskin}(2026)}]{iskin25c}%
  \BibitemOpen
  \bibfield  {author} {\bibinfo {author} {\bibfnamefont {M.}~\bibnamefont {Iskin}},\ }\bibfield  {title} {\bibinfo {title} {Quenched disorder and the {BCS-BEC} crossover in the {Hubbard} model},\ }\href {https://doi.org/10.1103/z2qj-rc9w} {\bibfield  {journal} {\bibinfo  {journal} {Phys. Rev. A}\ }\textbf {\bibinfo {volume} {113}},\ \bibinfo {pages} {013312} (\bibinfo {year} {2026})}\BibitemShut {NoStop}%
\bibitem [{\citenamefont {Che}\ \emph {et~al.}(2017)\citenamefont {Che}, \citenamefont {Zhang}, \citenamefont {Wang},\ and\ \citenamefont {Chen}}]{che17}%
  \BibitemOpen
  \bibfield  {author} {\bibinfo {author} {\bibfnamefont {Y.}~\bibnamefont {Che}}, \bibinfo {author} {\bibfnamefont {L.}~\bibnamefont {Zhang}}, \bibinfo {author} {\bibfnamefont {J.}~\bibnamefont {Wang}},\ and\ \bibinfo {author} {\bibfnamefont {Q.}~\bibnamefont {Chen}},\ }\bibfield  {title} {\bibinfo {title} {Impurity effects on {BCS-BEC} crossover in ultracold atomic {Fermi} gases},\ }\href {https://doi.org/10.1103/PhysRevB.95.014504} {\bibfield  {journal} {\bibinfo  {journal} {Phys. Rev. B}\ }\textbf {\bibinfo {volume} {95}},\ \bibinfo {pages} {014504} (\bibinfo {year} {2017})}\BibitemShut {NoStop}%
\bibitem [{\citenamefont {S\'a~de Melo}\ \emph {et~al.}(1993)\citenamefont {S\'a~de Melo}, \citenamefont {Randeria},\ and\ \citenamefont {Engelbrecht}}]{sademelo93}%
  \BibitemOpen
  \bibfield  {author} {\bibinfo {author} {\bibfnamefont {C.~A.~R.}\ \bibnamefont {S\'a~de Melo}}, \bibinfo {author} {\bibfnamefont {M.}~\bibnamefont {Randeria}},\ and\ \bibinfo {author} {\bibfnamefont {J.~R.}\ \bibnamefont {Engelbrecht}},\ }\bibfield  {title} {\bibinfo {title} {Crossover from {BCS} to {B}ose superconductivity: Transition temperature and time-dependent {G}inzburg-{L}andau theory},\ }\href {https://doi.org/10.1103/PhysRevLett.71.3202} {\bibfield  {journal} {\bibinfo  {journal} {Phys. Rev. Lett.}\ }\textbf {\bibinfo {volume} {71}},\ \bibinfo {pages} {3202} (\bibinfo {year} {1993})}\BibitemShut {NoStop}%
\bibitem [{\citenamefont {Engelbrecht}\ \emph {et~al.}(1997)\citenamefont {Engelbrecht}, \citenamefont {Randeria},\ and\ \citenamefont {S\'a~de Melo}}]{engelbrecht97}%
  \BibitemOpen
  \bibfield  {author} {\bibinfo {author} {\bibfnamefont {J.~R.}\ \bibnamefont {Engelbrecht}}, \bibinfo {author} {\bibfnamefont {M.}~\bibnamefont {Randeria}},\ and\ \bibinfo {author} {\bibfnamefont {C.~A.~R.}\ \bibnamefont {S\'a~de Melo}},\ }\bibfield  {title} {\bibinfo {title} {{BCS} to {B}ose crossover: Broken-symmetry state},\ }\href {https://doi.org/10.1103/PhysRevB.55.15153} {\bibfield  {journal} {\bibinfo  {journal} {Phys. Rev. B}\ }\textbf {\bibinfo {volume} {55}},\ \bibinfo {pages} {15153} (\bibinfo {year} {1997})}\BibitemShut {NoStop}%
\bibitem [{\citenamefont {Taylor}\ \emph {et~al.}(2006)\citenamefont {Taylor}, \citenamefont {Griffin}, \citenamefont {Fukushima},\ and\ \citenamefont {Ohashi}}]{taylor06}%
  \BibitemOpen
  \bibfield  {author} {\bibinfo {author} {\bibfnamefont {E.}~\bibnamefont {Taylor}}, \bibinfo {author} {\bibfnamefont {A.}~\bibnamefont {Griffin}}, \bibinfo {author} {\bibfnamefont {N.}~\bibnamefont {Fukushima}},\ and\ \bibinfo {author} {\bibfnamefont {Y.}~\bibnamefont {Ohashi}},\ }\bibfield  {title} {\bibinfo {title} {Pairing fluctuations and the superfluid density through the {BCS-BEC} crossover},\ }\href {https://doi.org/10.1103/PhysRevA.74.063626} {\bibfield  {journal} {\bibinfo  {journal} {Phys. Rev. A}\ }\textbf {\bibinfo {volume} {74}},\ \bibinfo {pages} {063626} (\bibinfo {year} {2006})}\BibitemShut {NoStop}%
\bibitem [{\citenamefont {Diener}\ \emph {et~al.}(2008)\citenamefont {Diener}, \citenamefont {Sensarma},\ and\ \citenamefont {Randeria}}]{diener08}%
  \BibitemOpen
  \bibfield  {author} {\bibinfo {author} {\bibfnamefont {R.~B.}\ \bibnamefont {Diener}}, \bibinfo {author} {\bibfnamefont {R.}~\bibnamefont {Sensarma}},\ and\ \bibinfo {author} {\bibfnamefont {M.}~\bibnamefont {Randeria}},\ }\bibfield  {title} {\bibinfo {title} {Quantum fluctuations in the superfluid state of the {BCS-BEC} crossover},\ }\href {https://doi.org/10.1103/PhysRevA.77.023626} {\bibfield  {journal} {\bibinfo  {journal} {Phys. Rev. A}\ }\textbf {\bibinfo {volume} {77}},\ \bibinfo {pages} {023626} (\bibinfo {year} {2008})}\BibitemShut {NoStop}%
\bibitem [{\citenamefont {Nozieres}\ and\ \citenamefont {Schmitt-Rink}(1985)}]{nsr85}%
  \BibitemOpen
  \bibfield  {author} {\bibinfo {author} {\bibfnamefont {P.}~\bibnamefont {Nozieres}}\ and\ \bibinfo {author} {\bibfnamefont {S.}~\bibnamefont {Schmitt-Rink}},\ }\bibfield  {title} {\bibinfo {title} {Bose condensation in an attractive fermion gas: From weak to strong coupling superconductivity},\ }\href {https://doi.org/10.1007/BF00683774} {\bibfield  {journal} {\bibinfo  {journal} {J. Low Temp. Phys.}\ }\textbf {\bibinfo {volume} {59}},\ \bibinfo {pages} {195} (\bibinfo {year} {1985})}\BibitemShut {NoStop}%
\bibitem [{Note1()}]{Note1}%
  \BibitemOpen
  \bibinfo {note} {We are grateful to the anonymous referee for pointing out that each of our disorder contributions can be naturally associated with such intuitive Feynman diagrams}\BibitemShut {NoStop}%
\bibitem [{\citenamefont {Abrikosov}\ \emph {et~al.}(1975)\citenamefont {Abrikosov}, \citenamefont {Dzyaloshinskii}, \citenamefont {Gorkov},\ and\ \citenamefont {Silverman}}]{abrikosov}%
  \BibitemOpen
  \bibfield  {author} {\bibinfo {author} {\bibfnamefont {A.~A.}\ \bibnamefont {Abrikosov}}, \bibinfo {author} {\bibfnamefont {I.}~\bibnamefont {Dzyaloshinskii}}, \bibinfo {author} {\bibfnamefont {L.~P.}\ \bibnamefont {Gorkov}},\ and\ \bibinfo {author} {\bibfnamefont {R.~A.}\ \bibnamefont {Silverman}},\ }\href {https://cds.cern.ch/record/107441} {\emph {\bibinfo {title} {{Methods of quantum field theory in statistical physics}}}}\ (\bibinfo  {publisher} {Dover},\ \bibinfo {address} {New York, NY},\ \bibinfo {year} {1975})\BibitemShut {NoStop}%
\bibitem [{\citenamefont {Lopatin}\ and\ \citenamefont {Vinokur}(2002)}]{lopatin02}%
  \BibitemOpen
  \bibfield  {author} {\bibinfo {author} {\bibfnamefont {A.~V.}\ \bibnamefont {Lopatin}}\ and\ \bibinfo {author} {\bibfnamefont {V.~M.}\ \bibnamefont {Vinokur}},\ }\bibfield  {title} {\bibinfo {title} {Thermodynamics of the superfluid dilute {Bose} gas with disorder},\ }\href {https://doi.org/10.1103/PhysRevLett.88.235503} {\bibfield  {journal} {\bibinfo  {journal} {Phys. Rev. Lett.}\ }\textbf {\bibinfo {volume} {88}},\ \bibinfo {pages} {235503} (\bibinfo {year} {2002})}\BibitemShut {NoStop}%
\bibitem [{\citenamefont {Iskin}(2025)}]{iskin25b}%
  \BibitemOpen
  \bibfield  {author} {\bibinfo {author} {\bibfnamefont {M.}~\bibnamefont {Iskin}},\ }\bibfield  {title} {\bibinfo {title} {Structure factors and quantum geometry in multiband {BCS} superconductors},\ }\href {https://doi.org/10.1103/67gs-51xd} {\bibfield  {journal} {\bibinfo  {journal} {Phys. Rev. B}\ }\textbf {\bibinfo {volume} {112}},\ \bibinfo {pages} {014517} (\bibinfo {year} {2025})}\BibitemShut {NoStop}%
\bibitem [{Note2()}]{Note2}%
  \BibitemOpen
  \bibinfo {note} {For the three-dimensional continuum model in the BEC limit, the condition $|\mu | \gg \Lambda ^2/(2m)$ guarantees that the characteristic pair size $\xi _{\protect \mathrm {pair}}$ is much smaller than the disorder correlation length~\cite {palestini13}. Since $\xi _{\protect \mathrm {pair}} \propto a_s$ and $|\mu | \simeq 1/(2m a_s^2)$ in this regime, the required separation of length scales is automatically satisfied. An analogous argument applies to the lattice model, where in the strong-coupling limit $U/t \gg 1$ one has $|\mu | \simeq U/2$ and $\varepsilon _{\max } = 12t$, while the pair size scales as $\xi _{\protect \mathrm {pair}} \propto a\protect \, t/U$. Under these conditions, the composite bosons respond to the disorder potential as effectively point-like particles}\BibitemShut {NoStop}%
\bibitem [{\citenamefont {{T{\"o}rm{\"a}}}\ \emph {et~al.}(2022)\citenamefont {{T{\"o}rm{\"a}}}, \citenamefont {{Peotta}},\ and\ \citenamefont {{Bernevig}}}]{torma22}%
  \BibitemOpen
  \bibfield  {author} {\bibinfo {author} {\bibfnamefont {P.}~\bibnamefont {{T{\"o}rm{\"a}}}}, \bibinfo {author} {\bibfnamefont {S.}~\bibnamefont {{Peotta}}},\ and\ \bibinfo {author} {\bibfnamefont {B.~A.}\ \bibnamefont {{Bernevig}}},\ }\bibfield  {title} {\bibinfo {title} {{Superconductivity, superfluidity and quantum geometry in twisted multilayer systems}},\ }\href {https://doi.org/10.1038/s42254-022-00466-y} {\bibfield  {journal} {\bibinfo  {journal} {Nature Reviews Physics}\ }\textbf {\bibinfo {volume} {4}},\ \bibinfo {pages} {528} (\bibinfo {year} {2022})}\BibitemShut {NoStop}%
\bibitem [{\citenamefont {Yu}\ \emph {et~al.}(2025)\citenamefont {Yu}, \citenamefont {Bernevig}, \citenamefont {Queiroz}, \citenamefont {Rossi}, \citenamefont {T{\"o}rm{\"a}},\ and\ \citenamefont {Yang}}]{yu24}%
  \BibitemOpen
  \bibfield  {author} {\bibinfo {author} {\bibfnamefont {J.}~\bibnamefont {Yu}}, \bibinfo {author} {\bibfnamefont {B.~A.}\ \bibnamefont {Bernevig}}, \bibinfo {author} {\bibfnamefont {R.}~\bibnamefont {Queiroz}}, \bibinfo {author} {\bibfnamefont {E.}~\bibnamefont {Rossi}}, \bibinfo {author} {\bibfnamefont {P.}~\bibnamefont {T{\"o}rm{\"a}}},\ and\ \bibinfo {author} {\bibfnamefont {B.-J.}\ \bibnamefont {Yang}},\ }\bibfield  {title} {\bibinfo {title} {Quantum geometry in quantum materials},\ }\href@noop {} {\bibfield  {journal} {\bibinfo  {journal} {npj Quantum Materials}\ }\textbf {\bibinfo {volume} {10}},\ \bibinfo {pages} {101} (\bibinfo {year} {2025})}\BibitemShut {NoStop}%
\bibitem [{\citenamefont {Jiang}\ \emph {et~al.}(2025)\citenamefont {Jiang}, \citenamefont {Holder},\ and\ \citenamefont {Yan}}]{jiang25}%
  \BibitemOpen
  \bibfield  {author} {\bibinfo {author} {\bibfnamefont {Y.}~\bibnamefont {Jiang}}, \bibinfo {author} {\bibfnamefont {T.}~\bibnamefont {Holder}},\ and\ \bibinfo {author} {\bibfnamefont {B.}~\bibnamefont {Yan}},\ }\bibfield  {title} {\bibinfo {title} {Revealing quantum geometry in nonlinear quantum materials},\ }\href@noop {} {\bibfield  {journal} {\bibinfo  {journal} {Reports on Progress in Physics}\ } (\bibinfo {year} {2025})}\BibitemShut {NoStop}%
\end{thebibliography}%

\end{document}